\documentclass[amsmath,amssymb,floatfix,prb,epsf,showpacs,twocolumn]{revtex4-1}
\usepackage{amsmath,amssymb,natbib,bm,graphicx,url,epsfig}
\usepackage[ansinew]{inputenc}
\usepackage[bookmarks=true]{hyperref}
\usepackage{bm}
\usepackage{color}
\usepackage[justification=raggedright,format=plain,indention=.3cm,font=small]{caption}
\usepackage{subcaption}

\newcommand{\be}{\begin{equation}}
\newcommand{\ee}{\end{equation}}
\newcommand{\bes}{\begin{equation}\begin{split}}
\newcommand{\ees}{\end{split}\end{equation}}
\newcommand{\bea}{\begin{eqnarray}}
\newcommand{\eea}{\end{eqnarray}}

\def\beq{\begin{equation}}
\def\eeq{\end{equation}}
\def\bea{\begin{eqnarray}}
\def\eea{\end{eqnarray}}

\begin{document}

\title{Statistical mechanics of Coulomb gases as quantum theory on Riemann surfaces\\
{\em \small Dedicated to the memory of Professor Anatoliy Larkin} }
\author{Tobias Gulden$^1$, Michael Janas$^1$, Peter Koroteev$^{1,2}$, and Alex Kamenev$^{1,3}$}

\affiliation{$^1$Department of Physics, University of Minnesota, Minneapolis, MN 55455, USA}
\affiliation{$^2$Perimeter Institute for Theoretical Physics, ON N2L2Y5, Canada}
\affiliation{$^3$William I. Fine Theoretical Physics Institute, University of Minnesota, Minneapolis, MN 55455, USA}

%\date{\currenttime \today}
\date{\today}
\vspace{0.1cm}

\begin{abstract}
Statistical mechanics of 1D multivalent Coulomb gas may be mapped onto non-Hermitian quantum mechanics. We use
this example to develop instanton calculus on Riemann surfaces. Borrowing from the formalism developed in the context of Seiberg-Witten duality, we treat momentum and coordinate as complex variables. Constant energy manifolds are given
by  Riemann surfaces of genus $g\geq 1$. The actions along principal cycles on these surfaces obey ODE
in the moduli space of the Riemann surface known as Picard-Fuchs equation. We derive and solve Picard-Fuchs equations
for Coulomb gases of various charge content. Analysis of monodromies of these solutions
around their singular points yields semiclassical  spectra  as well as  instanton effects such as Bloch bandwidth. Both are shown to be in perfect agreement with numerical simulations.
\end{abstract}
%\pacs{XX.XX.Pm, YY.YY.Fg, ZZ.ZZ.Tg}

\maketitle

\section{Introduction}

One of the very last works of Anatoliy Larkin\cite{Larkin} was devoted to transport through ion channels of biological
membranes. An ion channel may be roughly viewed as a cylindrical water-filled tube surrounded by a lipid membrane. Its typical
radius $a\approx 6\AA$ is much smaller than its length $L\approx 120\AA$. The important observation with far reaching consequences,
made in Ref.~[\onlinecite{Larkin}], is that the dielectric constant of water $\epsilon_{\rm water}\approx 80$ is significantly larger than that
of the surrounding lipid membrane $\epsilon_{\rm lipid}\approx 2$. This defines a new length scale $\xi \approx a\sqrt{\epsilon_{\rm water}/\epsilon_{\rm lipid}}\ln(\epsilon_{\rm water}/\epsilon_{\rm lipid}) \approx 140\AA $ over which the electric field stays inside the channel and does not escape
into the surrounding media. Since $\xi\gtrsim L$, the ions inside the channel interact essentially through the 1D Coulomb potential
$U(x_1-x_2) \approx eE_0|x_1-x_2|$, where $E_0=2e/a^2\epsilon_{\rm water}$ is a discontinuity of the electric field created by a unit charge. This fact dictates a significant energy barrier $U(L/4) \approx 4 k_B T_{\rm room}$
for moving a single ion through the channel. If indeed present, such a barrier would essentially impede ion transport, preventing the channel
from performing its biological functions.

Nature removes such Coulomb blocking by screening. A moving ion is screened either by mobile ions of dissociated salt\cite{Larkin}, or by immobilized charged radicals attached to the walls of the channel\cite{ZhangPRE,ZhangPRL,MacKinnon,Doyle,Roux,Allen,Chung,Berezhkovskii}. Nevertheless, due to the peculiar nature of the long-range 1D Coulomb potential, the transport barrier proportional to the channel length $L$ is always present. Its
magnitude, though, is typically suppressed\cite{Larkin} down to about $k_BT_{\rm room}$, allowing
for a relatively unimpeded transport of ions. These considerations call for development of a transport theory of 1D Coulomb gases. Following the celebrated mapping of 1D statistical mechanics onto an effective quantum mechanics, pioneered by Edwards and Lenard \cite{EdwardsLenard} and Vaks, Larkin and Pikin\cite{VaksLarkin}, reference~[\onlinecite{Larkin}] mapped the problem onto quantum mechanics of a cosine potential (we briefly review this mapping in Sec. \ref{Background}). The ground state
energy of such quantum mechanics is exactly the equilibrium pressure in the Coulomb plasma. Moreover the width of the lowest Bloch band is a specific energy barrier for ion transport through the channel.

It is instructive to notice that  $2\alpha \cos\theta =\alpha(e^{i\theta} + e^{-i\theta})$ potential describes  a mixture of positive, $e^{i\theta}$, and negative, $e^{-i\theta}$, {\em monovalent} ions with concentration $\alpha$. One may also  consider a situation when the channel
is filled with a solution of dissociated {\em multivalent} salt, such as e.g. divalent CaCl$_2$ or trivalent AlCl$_3$. In these cases the corresponding 1D statistical mechanics is mapped onto the quantum problem with a {\em non-Hermitian} potential such as $ \alpha({1\over 2} e^{2i\theta} + e^{-i\theta})$ or $\,\alpha({1\over 3}e^{3i\theta} + e^{-i\theta})$ \cite{EdwardsLenard,ZhangPRE}. The present paper is devoted to efficient mathematical methods of treating non-Hermitian quantum mechanics of this sort.

Our particular focus here is on a {\em semiclassical} treatment, applicable in the regime of sufficiently large salt concentration $\alpha$.
In its framework the energy spectrum (thus the pressure) is determined by the Bohr-Sommerfeld quantization condition for the action of classical periodic orbits. On the other hand, the bandwidth (and thus the transport barrier) is given by the exponentiated  action accumulated on the {\em instanton}
trajectory, running through the classically forbidden  part of the phase space. The traditional techniques of Hermitian quantum mechanics call for finding classical and instanton trajectories by solving equations of motion in real and imaginary time and evaluating corresponding actions. This route can't be straightforwardly applied to non-Hermitian quantum problems arising in the context of multi-valent Coulomb gases.
Even leaving aside the technical difficulties of solving complex equations of motion, there are conceptual difficulties with identifying periodic orbits as well as the meaning of classically allowed vs. forbidden regions and with the imaginary time procedure.

In this paper we borrow from the algebraic topology methods developed in the past decades in the context of the Seiberg-Witten solution \cite{Seiberg:1994rs,Seiberg:1994aj} and its applications to integrable systems\cite{Donagi:1995cf,Gorsky:1995zq,Gorsky:2010} (and many follow-up contributions). The central idea is to consider both coordinate $\theta$ and corresponding canonical momentum $p$ as {\em complex} variables. This leads to  {\em four}-dimensional (4D) phase space. Then (complex) energy conservation restricts the trajectories to live on 2D Riemann surfaces embedded into 4D phase space. The dynamics of the system are essentially determined by the topology, i.e. genus $g$, of such Riemann surfaces.
We show that e.g. mono- and divalent gases are described by tori, while trivalent and 4-valent  lead to genus-$2$ surfaces, etc.
The Cauchy theorem and the resulting freedom to deform the integration contour in the complex space allows  to avoid finding specific solutions
of the equations of motion. Instead one identifies the homology cycles on the Riemann surface and finds the corresponding action integrals, which depend only on the topology of the cycles and not on their specific shape. For example, the cosine potential of monovalent gas leads to a torus, which obviously has two topologically distinct cycles, Fig.~\ref{fig:DegenerateTorus}. The two turn out to be related to classical and instanton actions  correspondingly. The genus $g\geq 1$ Riemann surfaces admit $2g$ topologically distinct cycles. Below we identify and explain the meaning of the corresponding action integrals.

The shape of the specific Riemann surface depends on the parameters of the problem, e.g. salt concentration $\alpha$ in our case. Such parameters are called moduli of the Riemann surface. It turns out that the action integrals, being functions of the moduli, satisfy closed ordinary differential equation (ODE) of the order $2g$, known as the Picard-Fuchs equation. The actions may be found as solutions of this ODE in the moduli space, rather than performing integrations over cycles on the surface. Below we derive and solve Picard-Fuchs equations for several (positive, negative) ionic charge combinations, such as genus $g=1$ cases $(1,1), (2,1)$ and genus $g=2$ cases $(3,1), (3,2), (4,1)$.
We then discuss how to connect the principal classical actions with the spectra of the corresponding quantum problem. The key observation is that in the moduli space the actions exhibit a few isolated branching points. Going around such a branching point transforms the actions into
their linear combinations -- the $Sp(2g,\mathbb{Z})$ monodromy transformation. The invariance of  quantum observables under monodromy transformations dictates Bohr-Sommerfeld quantization for one of the principal classical actions. The remaining actions may be identified with the instanton processes, related to e.g. Bloch bandwidth.

Statistical mechanics of 1D Coulomb gases may seem to be an isolated problem, not worthy of developing an extensive mathematical apparatus.
Our goal here is to use it as a test-drive example, grounded into a well-posed physics problem, to develop a machinery applicable in other setups. Recently the so-called ${\cal PT}$ symmetric non-Hermitian quantum mechanics attracted a lot of attention for its application in active optics\cite{Optics} and open quantum systems\cite{OpenQuantumSystems}, as well as in the description of antiferromagnetic lattices\cite{Antiferromagnets} and calculating energy states in larger molecules\cite{Diatomics}. Our examples also belong to the class of ${\cal PT}$ symmetric problems. It seems likely that the methods developed here may be applied to advance analytical understanding of a broader class of ${\cal PT}$ symmetric quantum mechanics. Another context, where complexified quantum mechanics was proven to be extremely useful, is dynamics of large molecular spins\cite{StoneGarg,KececiogluGarg}. Indeed functional integral representation of the spin dynamics leads naturally to the Hamiltonian formulation,  where the projective coordinates $(z,\bar z)$ on the sphere play the role of the canonical pair\cite{AltlandSimons}.
It was realized\cite{StoneGarg,KececiogluGarg} that to find instanton trajectories one has to consider $z$ and $\bar z$ as independent complex
variables, thus expanding the dynamics into 4D phase space. The Riemann geometry methods seem to be well-suited to advance this subject as well.

This paper is organized as follows:  in section \ref{Background} we outline the relation between 1D multivalent Coulomb gases and
non-Hermitian quantum mechanics and discuss general symmetries of the latter. In section \ref{Numerics} we summarize major numerical
observations regarding complex spectra and band-structure for the family of Hamiltonians considered here. In section \ref{sec:monovalent} we illustrate the
machinery of algebraic geometry on Riemann surfaces for the familiar Hermitian cosine potential quantum mechanics, which corresponds to the monovalent $(1,1)$ gas. Here we introduce complexified phase space and Riemann
torii  of constant energy; we then derive, solve and analyze solutions of the Picard-Fuchs equations. In section \ref{sec:divalent} we apply the developed methods for the divalent $(2,1)$ Coulomb gas, which is also described by a genus-1 torus. In section \ref{sec:trivalent} we extend the method for genus-2 example of trivalent $(3,1)$ gas, which exhibits some qualitatively new features. The $(3,2)$ and $(4,1)$ gases are briefly tackled in section \ref{sec:32-41}. In section \ref{sec:SW} we outline connections to Seiberg-Witten theory. We conclude with a brief discussion of the results in section \ref{sec:discussion}.

\section{Mapping of Coulomb gases onto quantum mechanics}
\label{Background}

Consider a 1D gas of cations with charge $n_1e$ and anions with charge $-n_2e$, where $(n_1,n_2)$ are positive integers. By Gauss's theorem, the electric field at a
distance $x$ larger than the radius of the channel $a$ from a unit charge is  $E_0=2e/a^2\epsilon_{\rm water}$. At the location of a charge $n_{1,2}$ the electric field exhibits a discontinuity $\pm 2E_0 n_{1,2}$. Since all charges are integers the field is conserved modulo $2E_0$ along the channel. This allows us to define the order parameter\cite{Larkin,ZhangPRL} $q=E(x) (\!\mod\, 2E_0)$, which acts like an
effective boundary charge $\pm q$ at the two ends of the channel. The Poisson equation in 1D reads $\nabla^2\phi=-2E_0\delta(x)$, leading to  1D Coulomb potential  $\phi(x)=-E_0|x|$. The potential energy of the gas is thus
\begin{equation}
U=-\frac{eE_0}{2}\sum_{i,j}\sigma_i\sigma_j|x_i-x_j|\,,
\label{eq:potential}
\end{equation}
where $\sigma_j$ is the charge $n_1$ or $-n_2$ of an ion at the position $x_j$ and we omit the $\pm q$ boundary charges for brevity. Our goal is to evaluate the grand canonical partition function of the gas in the channel of length $L$
\begin{equation}
\mathcal{Z}_L=\!\!\! \sum_{N_1,N_2=0}^{\infty}\frac{f_1^{N_1}f_2^{N_2}}{N_1!N_2!}\prod_{i=1}^{N_1}\int_0^Ldx_i\prod_{j=1}^{N_2}\int_0^Ldx_j\,\, e^{-U/k_BT}\, ,
\label{eq:partition}
\end{equation}
where $f_{1,2}$ are fugacities of the two charge species. One can now introduce the charge density employing a delta-function
$\delta[\rho(x)-\sum_j\sigma_j\delta(x-x_j)]$. The delta-function is elevated in the exponent with the help of the auxiliary field $\theta(x)$. This  procedure decouples all $x_j$ integrals\cite{Larkin}, bringing them to the form $\sum_N [f\int\! dx\, e^{i\sigma \theta(x)}]^N/N!=\exp\{f
\int\! dx\, e^{i\sigma \theta(x)}\}$. The interaction potential (\ref{eq:potential}), being inverse of the 1D Laplace operator, leads to $\exp\{(T/eE_0)\int\! dx\,  \theta \partial_x^2 \theta\}$. As a result the partition function (\ref{eq:partition}) is identically written as the Feynman
path integral, in an ``imaginary time'' $x$,  for the quantum mechanics with the Hamiltonian
\begin{equation}
\hat{H}=(i\partial_\theta-q)^2-\left(\alpha_1 e^{in_1\theta}+\alpha_2 e^{-in_2\theta} \right)\, ,
\label{eq:Hamiltonian}
\end{equation}
where $\alpha_{1,2}=f_{1,2}k_BT/eE_0$ are dimensionless ion concentrations. Such  Feynman integral is the expectation value of the evolution
operator during ``time'' $L$, leading to
\begin{equation}
                                                                  \label{eq:partition1}
 \mathcal{Z}_L=\left\langle q\Big|{\cal X} e^{-\frac{eE_0}{k_BT}\int_0^L\! dx\, \hat{H} } \Big|q\right\rangle
 =\sum_m |\langle q|m\rangle|^2  e^{-\frac{eE_0L}{k_BT} \epsilon_m(q)},
\end{equation}
where ${\cal X}$ stands for $x$-ordered exponent.
Here $\epsilon_m(q)$ are eigenvalues of the effective Hamiltonian $\hat{H}$ and $|m\rangle =\psi_m(\theta)$ are its eigenvectors
in the Hilbert space of periodic functions $\psi_m(\theta)=\psi_m(\theta+2\pi)$, and finally the matrix elements are $ \langle q|m\rangle= \int_0^{2\pi}\!\! d\theta e^{-iq\theta} \psi_m(\theta)$. The boundary charge $q$ plays the role of the Bloch quasi-momentum and the spectrum is obviously periodic in $q$ with the unit period (reflecting the fact that the integer part of the boundary charge may be screened by mobile
ions and thus inconsequential).

The pressure of the Coulomb gas is its free energy per unit length
\begin{equation}
                                                         \label{eq:pressure}
P= k_B T \frac{\partial \ln \mathcal{Z}_L}{\partial L} \,  \stackrel{L\to \infty}{\longrightarrow}\, -eE_0 \epsilon_0(q)\,,
\end{equation}
where $\epsilon_0(q)$ is the eigenvalue with the smallest real part. In equilibrium the system minimizes its free energy by choosing
an appropriate boundary charge $q$. In all cases considered below the minimum appears to be a non-polarized state of the channel, i.e. $q=0$ (see however Refs.~[\onlinecite{ZhangPRE}] for exceptions to this rule). Adiabatic charge transfer through the channel is associated with the boundary charge $q$ sweeping through its full period. As a result, the (free) energy barrier for ion transport is
\beq
                                                        \label{eq:barrier}
U_0=eE_0 L \Delta_0\,,
\eeq
where $\Delta_0$ is the width of the lowest Bloch band. Therefore the ground state energy and the width of the lowest Bloch band of the Hamiltonian (\ref{eq:Hamiltonian}) determine thermodynamic and transport properties of the $(n_1,n_2)$ Coulomb gas. The rest of this paper is devoted to a semiclassical theory of the spectral properties of such Hamiltonians. We start by discussing some general symmetries of the non-Hermitian Hamiltonian (\ref{eq:Hamiltonian}).

%This means that knowledge of the band spectrum of $\mathcal{H}_q$ allows us to determine the transport behavior of ions through the channel.\\
%The group of Hamiltonians in equation (\ref{Hamiltonian}) acts on the space of periodic functions $\psi(\theta+2\pi)=\psi(\theta)$, where the period %can be taken smaller if $n_1$ and $n_2$
%have common prime factors. For $n_1=n_2$ (charges with equal valency) and $\alpha_1=\alpha_2$ (same concentration of charges, together with $n_1=n_2$ %this implies charge neutrality) this
%Hamiltonian reduces to the Mathieu-Hamiltonian $\mathcal{H}_q=-(i\partial_\theta-q)^2-\frac{2\alpha}{n}\cos(n\theta)$, which was widely studied in %literature \cite{MeixnerSchaefke,CUMS}
%and applied to this system \cite{ZhangPRE,Larkin,ZhangPRL}. For $n_1=n_2\neq1$ a redefinition of the coordinate $\theta$ allows us to reduce this to the %case $n_1=n_2=1$, but for other
%combinations of charge numbers or $\alpha_1\neq\alpha_2$ this procedure is not possible.

\subsection{${\cal PT}$ Symmetry}
\label{subsec:PT-sym}

Although the Hamiltonian (\ref{eq:Hamiltonian}) is non-Hermitian for $n_1\neq n_2$, it obeys $\mathcal{PT}$-symmetry \cite{Bender2002,Bender2003}.
Here the parity operator $\mathcal{P}$ acts as $\theta\to -\theta$, while the time-reversal operator $\mathcal{T}$ works as complex conjugation $i\to -i$. Clearly the two operations combined leave the Hamiltonian (\ref{eq:Hamiltonian}) unchanged. One may prove \cite{Bender2003,JoglekarBarnett} that all eigenvalues of $\mathcal{PT}$-symmetric Hamiltonians are either real, or appear in complex conjugated pairs. As shown below for positive values of concentrations $\alpha_{1,2}>0$ the lowest energy band $\epsilon_0(q)$ is entirely real, ensuring the positivity of the partition function. The higher bands $\epsilon_m(q)$ are in general complex. It is interesting to note that, for unphysical negative concentrations $\alpha_{1,2}<0$,  already the lowest band $\epsilon_0(q)$ is complex, making the free energy ill-defined.

%In section \ref{Isospectrality} we will show that every Hamiltonian $n_1\alpha_1-n_2\alpha_2\neq0$, or in
%other words a system which is not charge neutral, can be mapped onto a system which obeys charge neutrality, $n_1\alpha_1-n_2\alpha_2=0$. From that %we conclude that for $n_1=n_2$ the
%$\mathcal{PT}$-symmetry does not break and the spectrum $\epsilon_n(q)$ is entirely real. But for $n_1\neq n_2$ $\mathcal{H}_q$ can not be mapped onto a manifestly Hermitian Hamiltonian,

\subsection{Isospectrality}
\label{Isospectrality}

The spectrum of the Hamiltonian (\ref{eq:Hamiltonian}) is invariant under shift of the coordinate $\theta\to \theta +\theta_0$, where
$\theta_0$ is an arbitrary complex number.  Upon such transformation (preserving the periodic boundary conditions) the dimensionless concentrations $\alpha_{1,2}$ renormalize
as $\alpha_1\to \alpha_1 e^{in_1\theta_0}$ and $\alpha_2\to \alpha_2 e^{-in_2\theta_0}$. Notice that the combination $\alpha_1^{n_2}\alpha_2^{n_1}$ remains invariant. From here one concludes that the family of Hamiltonians (\ref{eq:Hamiltonian}) with
\begin{equation}
                                                            \label{eq:isospectral}
\alpha_1^{n_2}\alpha_2^{n_1} = {\rm const}
\end{equation}
are {\em isospectral}\cite{EdwardsLenard}. Thus without loss of generality, one may pick one representative from each isospectral
family. It is convenient to choose such a representative to manifestly enforce charge neutrality in the bulk reservoirs. To this end one takes $\alpha_1n_1=\alpha_2n_2=\alpha$, which brings the Hamiltonian (\ref{eq:Hamiltonian}) to the form
\begin{equation}
                                                                              \label{HamiltonianIso}
 \hat{H} =\alpha\left[ \hat p^2-  \left(\frac{1}{n_1}\,e^{in_1\theta}+\frac{1}{n_2}\, e^{-in_2\theta}\right)\right]\,,
\end{equation}
where  we have defined the momentum operator as
\begin{equation}
                                              \label{eq:p}
\hat p = \alpha^{-1/2}(-i\partial_\theta +q)\,; \quad\quad [\theta, \hat p]=i\alpha^{-1/2}\,.
\end{equation}
The commutation relation shows that $\alpha^{-1/2}$ plays the role of the effective Planck constant.
With the help of the isospectrality condition (\ref{eq:isospectral}), one may always choose a proper $\alpha$ such that the spectrum of Hamiltonian (\ref{HamiltonianIso}) is identical with that of a Hamiltonian with arbitrary $\alpha_{1,2}$. The physical reason for this symmetry is that the interior region of the long channel always preserves charge neutrality, allowing the edge regions
to screen charge imbalance of the reservoirs. Therefore, irrespective of the relative fugacities of cations and anions in the reservoirs, the thermodynamics of the long channel are equivalent to the one in contact with neutral reservoirs with an appropriate salt concentration $\alpha$.
Hereafter we restrict ourselves to the neutral Hamiltonian (\ref{HamiltonianIso}) with the single parameter $\alpha$.

\section{Numerical Analysis}
\label{Numerics}

In this section we discuss numerical simulation of the spectrum of the Hamiltonian (\ref{HamiltonianIso}). We focus on unequal charges $n_1\neq n_2$, since the case of $n_1=n_2$ reduces to the well-known Hermitian cosine potential\cite{CUMS,MeixnerSchaefke}. For unequal charges the Hamiltonian is non-Hermitian but $\mathcal{PT}$-symmetric, allowing for complex eigenvalues which appear in conjugated pairs\cite{Bender2003,JoglekarBarnett}.

Since the Hamiltonian $\hat{H}$ acts in the Hilbert space of periodic functions, one may choose the complete basis in the form
$\{e^{im\theta}\}_{m\in\mathbb{Z}}$. In this basis the Hamiltonian is represented by an infinite size {\em real} matrix\cite{ZhangPRE}
\begin{equation}
\hat{H}_{m,m'}\!=\!(m-q)^2\delta_{m,m'}-\alpha\!\left(\frac{1}{n_1}\delta_{m+n_1,m'}+\frac{1}{n_2}\delta_{m-n_2,m'}\!\right).
%\{\hat{H}\}_{n,n'}=(n-q)^2\delta_{n,n'}-\alpha\left(\frac{1}{n_1}\,\delta_{n+n_1,n'}+\frac{1}{n_2}\,\delta_{n-n_2,n'}\right).
\label{Hamiltonian_matrix}
\end{equation}
The boundary charge $q$ plays  the role of  quasi-momentum residing in the Brillouin zone $q\in[-\frac{1}{2},\frac{1}{2}]$. To numerically calculate the energy spectrum $\epsilon_m(q)$  we truncate the matrix at a large cutoff, after checking that a further increase in the matrix size does not change the low-energy spectrum. We left the boundary conditions ``open'', i.e. did not change the matrix elements near the cutoff, after verifying that different boundary conditions don't affect the result. It is easy to see that the matrix size should be $\gg \sqrt{\alpha}$ to accurately represent the low-energy spectrum. As an illustration we show the Hamiltonian cut to a $5\times5$ matrix for divalent $(2,1)$ gas:
\begin{displaymath}
%\{\mathcal{H}_q\}\to
\left(
\begin{array}{ccccc}
(-2-q)^2 & 0 & -\frac{\alpha}{2} & 0 & 0\\
-\alpha & (-1-q)^2 & 0 & -\frac{\alpha}{2} & 0\\
0 & -\alpha & (0-q)^2 & 0 & -\frac{\alpha}{2}\\
0 & 0 & -\alpha & (1-q)^2 & 0\\
0 & 0 & 0 & -\alpha & (2-q)^2
\end{array}\right)
\end{displaymath}

\begin{figure}[!h]
\raggedright
\hspace{0.03\textwidth}
  \begin{subfigure}{2.0cm}
  \includegraphics[width=6cm]{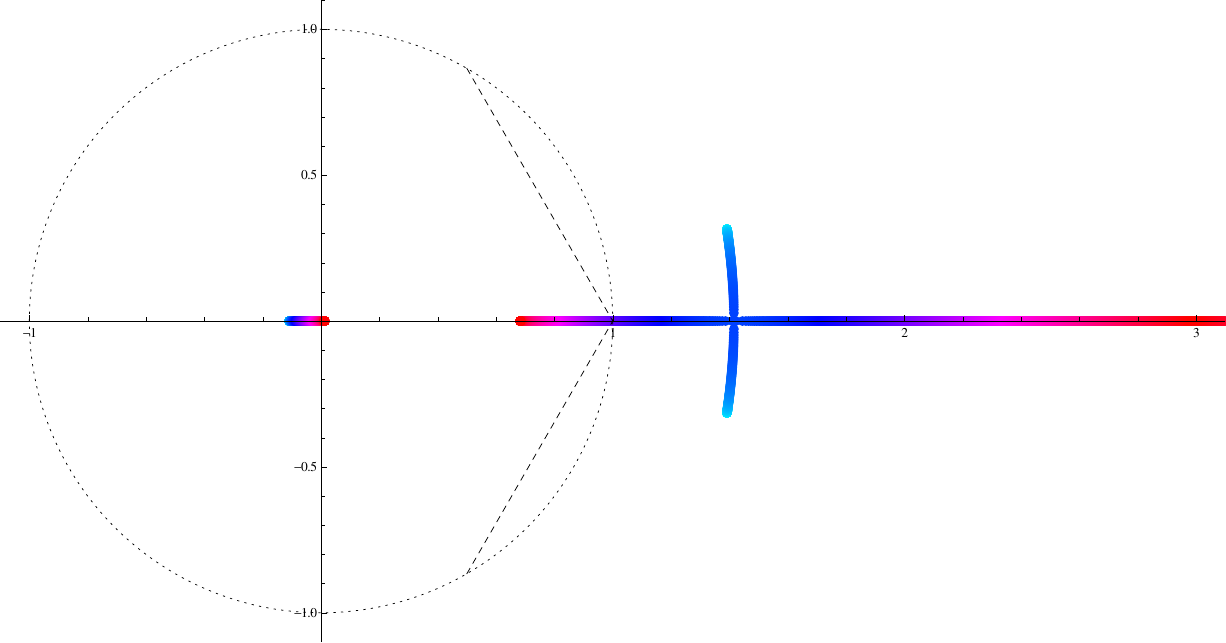}
  \caption{$\alpha=0.5$}
  \label{alpha0.5}
  \end{subfigure}

\hspace{0.03\textwidth}
  \begin{subfigure}{2.0cm}
  \includegraphics[width=6cm]{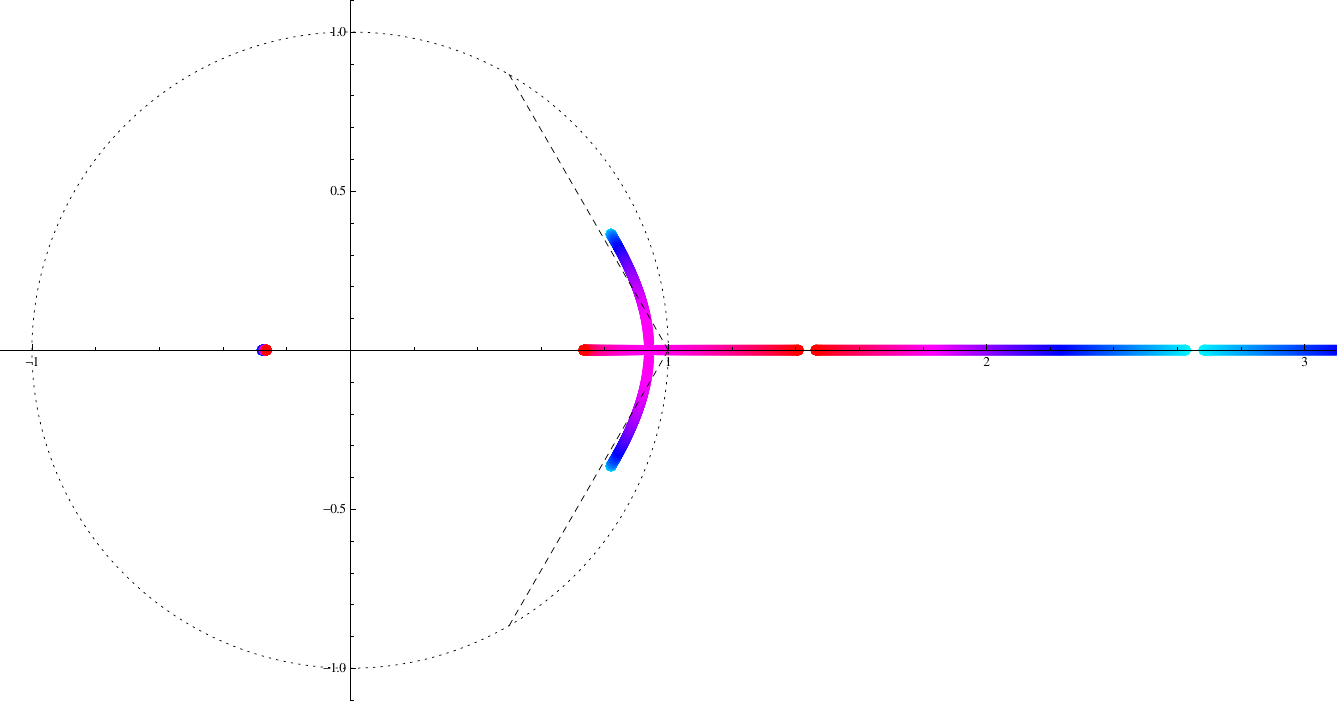}
  \caption{$\alpha=1$}
  \label{alpha1.0}
  \end{subfigure}

\hspace{0.03\textwidth}
  \begin{subfigure}{2.0cm}
  \includegraphics[width=6cm]{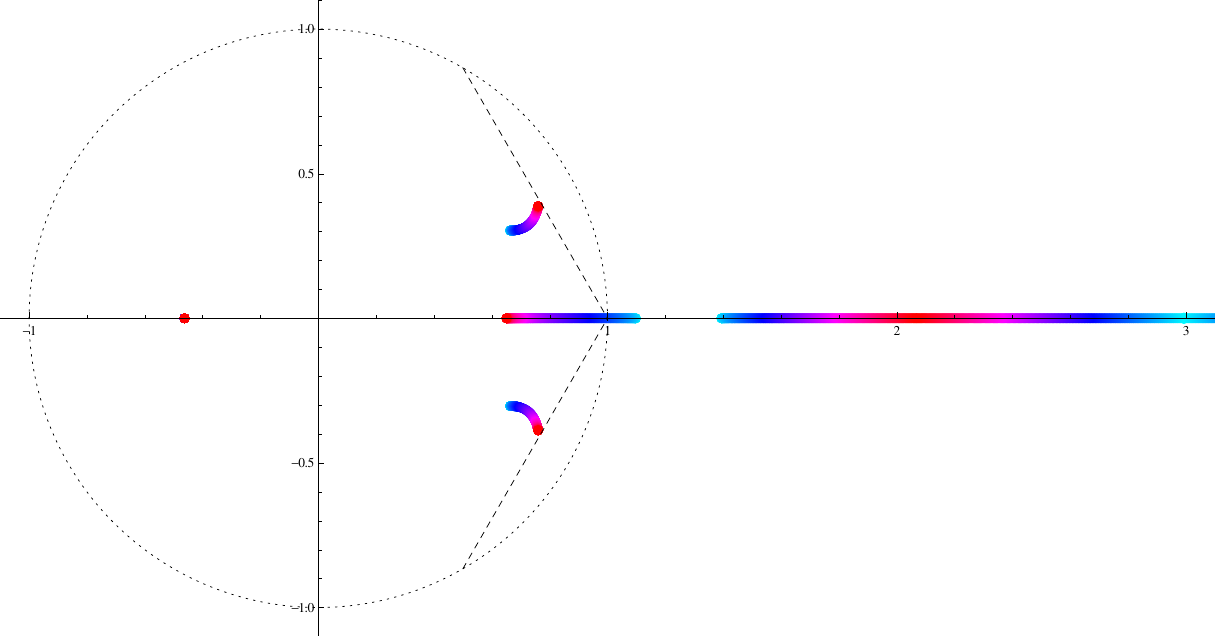}
  \caption{$\alpha=2$}
  \label{alpha2.0}
  \end{subfigure}

\hspace{0.03\textwidth}
  \begin{subfigure}{2.0cm}
  \includegraphics[width=6cm]{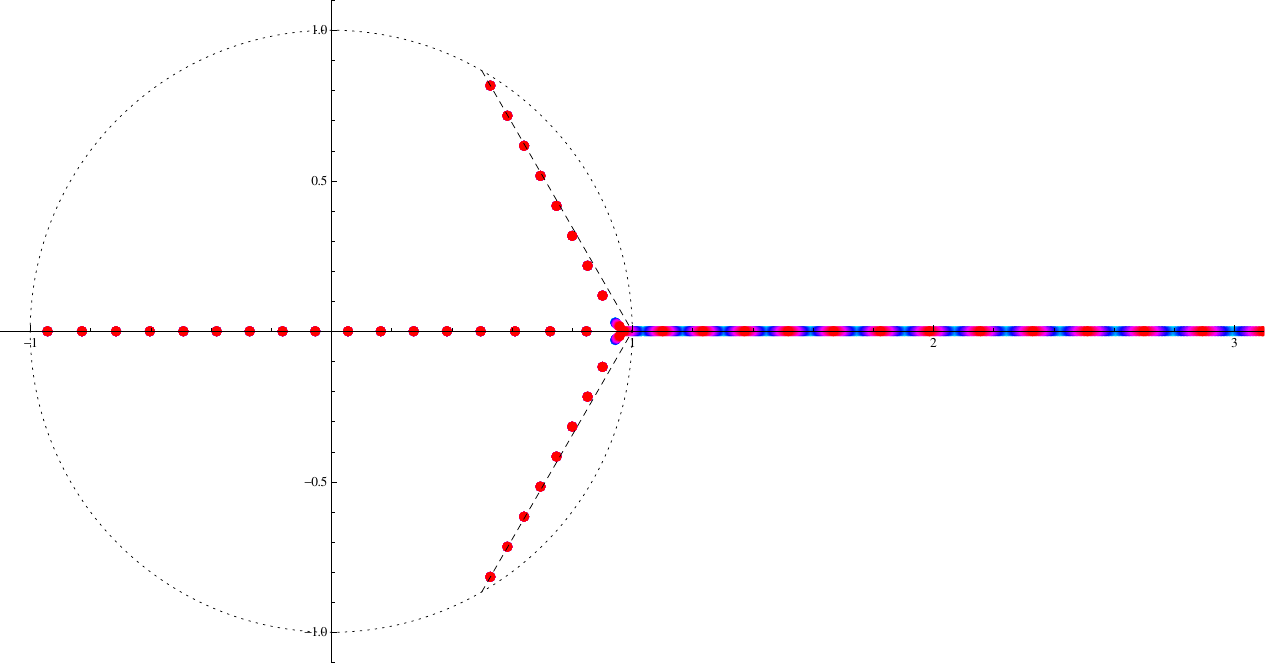}
  \caption{$\alpha=200$}
  \label{alpha200}
  \end{subfigure}
\caption{(Color online) Complex plane of normalized energy $u=2\epsilon_m(q)/3\alpha $ for  $(2,1)$ gas. The color corresponds to  different values of quasimomentum $q$; blue stands for $q=0$ and red for $q=\pm 1/2$. The dotted circle is $|u|=1$, the dashed lines connect between $u=1$ and $u=e^{\pm i \pi/3}$, indicating positions of the narrow complex bands in the limit of large $\alpha$.}
\label{EVinPlane_2/1}
\end{figure}

For  reasons which will become apparent below, it is convenient to present the spectrum $\epsilon$ on the complex plane of the
normalized energy $u$ defined as
\begin{equation}
                            \label{eq:u}
u=\frac{n_1n_2}{n_1+n_2}\,\frac{\epsilon}{\alpha}\,.
\end{equation}
For the divalent $(2,1)$ gas $u=2\epsilon/3\alpha$ and the corresponding  spectra are shown in Fig.~\ref{EVinPlane_2/1}. The spectrum consists of a sequence of complex Bloch bands. The number of narrow bands within the unit circle $|u|=1$ scales as $\sqrt{\alpha}$. They form three branches which terminate at $u=-1$ and $u=e^{\pm i \pi /3}$ and approximately line up along the lines connecting the termination points with the point $u=1$. We shall discuss the corresponding bandwidths below. Outside the unit circle the bands are wide and centered near the positive real axis of energy.
\begin{figure}
\includegraphics[width=8cm]{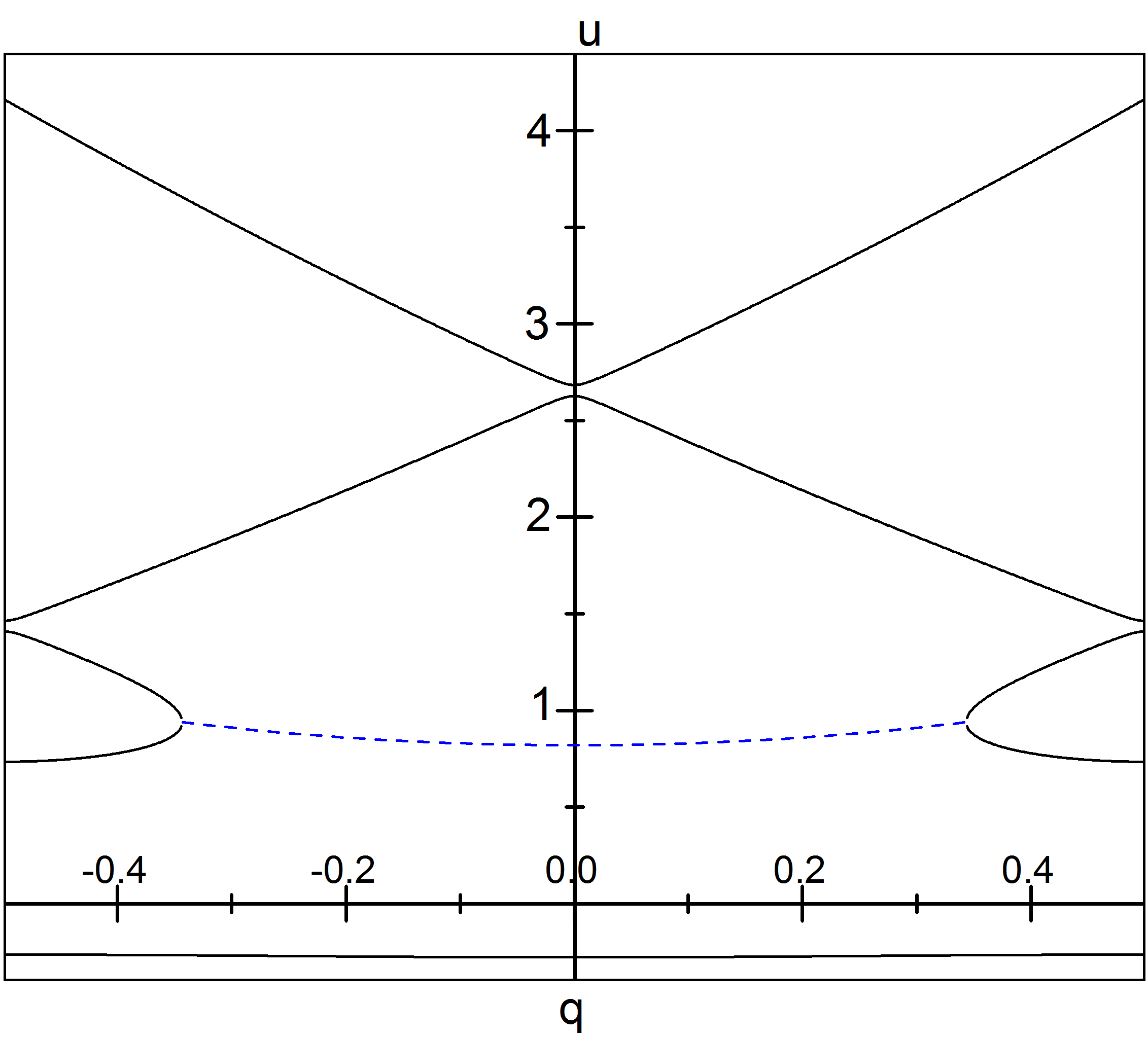}
\caption{(Color online) Band structure for  $(2,1)$ gas with $\alpha=1$, cf. Fig.~\ref{EVinPlane_2/1}b, vs. boundary charge (quasi- momentum) $q$. For the complex bands the real part of $\epsilon_m(q)$ is shown in dashed blue.}
\label{BandGraph}
\end{figure}

Figure \ref{BandGraph} shows the band structure in the first Brillouin zone $|q|<1/2$ for $\alpha=1$. Notice that the lowest Bloch band is purely real (this is always the case for $\alpha>0$), ensuring positive partition function (\ref{eq:partition1}) and real pressure (\ref{eq:pressure}). The next
{\em two} bands are complex. For $|q|<q_c\approx 0.36$ they exhibit opposite imaginary parts (not shown), but turn real at $|q|>q_c$. The next two bands are real, cf. Fig.~\ref{EVinPlane_2/1}b.  The higher bands form an alternating sequence of  two real and two complex bands. For larger values of $\alpha$ there is a sequence of entirely complex narrow bands, cf. Fig.~\ref{EVinPlane_2/1}d.

\begin{figure}[h!]
\raggedright
\hspace{0.03\textwidth}
  \begin{subfigure}{2.0cm}
  \includegraphics[width=6cm]{Images/ComplexEnergyPlane_2,1_200.pdf}
  \caption{$(2,1)$}
  \label{n_2,1}
  \end{subfigure}

\hspace{0.03\textwidth}
  \begin{subfigure}{2.0cm}
  \includegraphics[width=6cm]{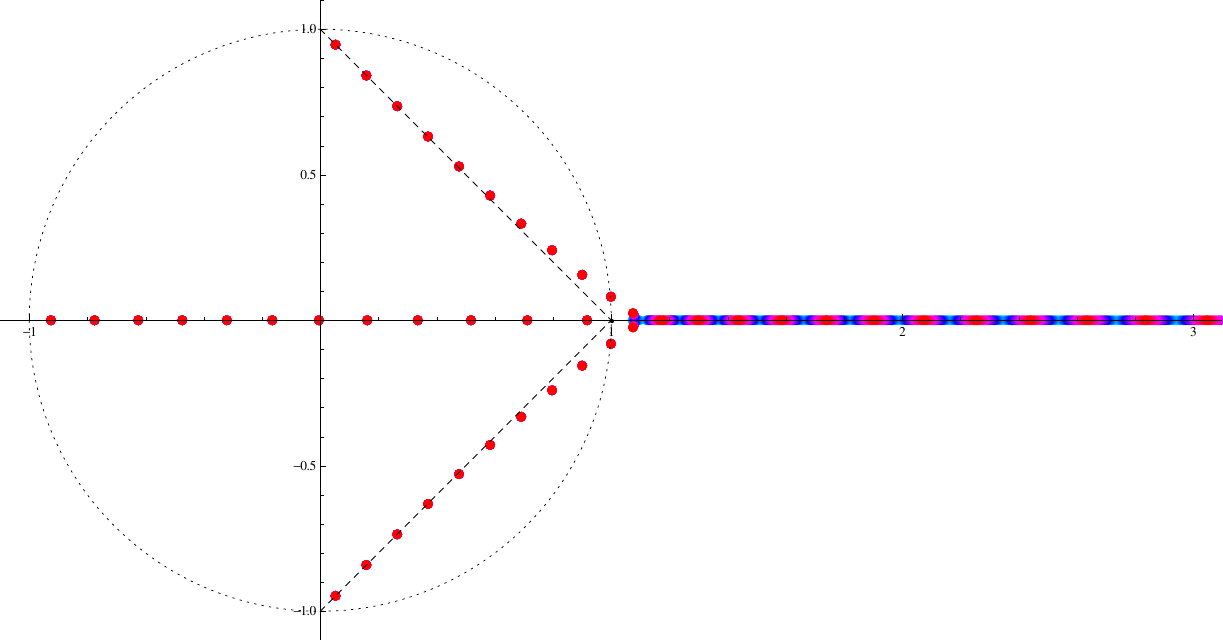}
  \caption{$(3,1)$}
  \label{n_3,1}
  \end{subfigure}

\hspace{0.03\textwidth}
  \begin{subfigure}{2.0cm}
  \includegraphics[width=6cm]{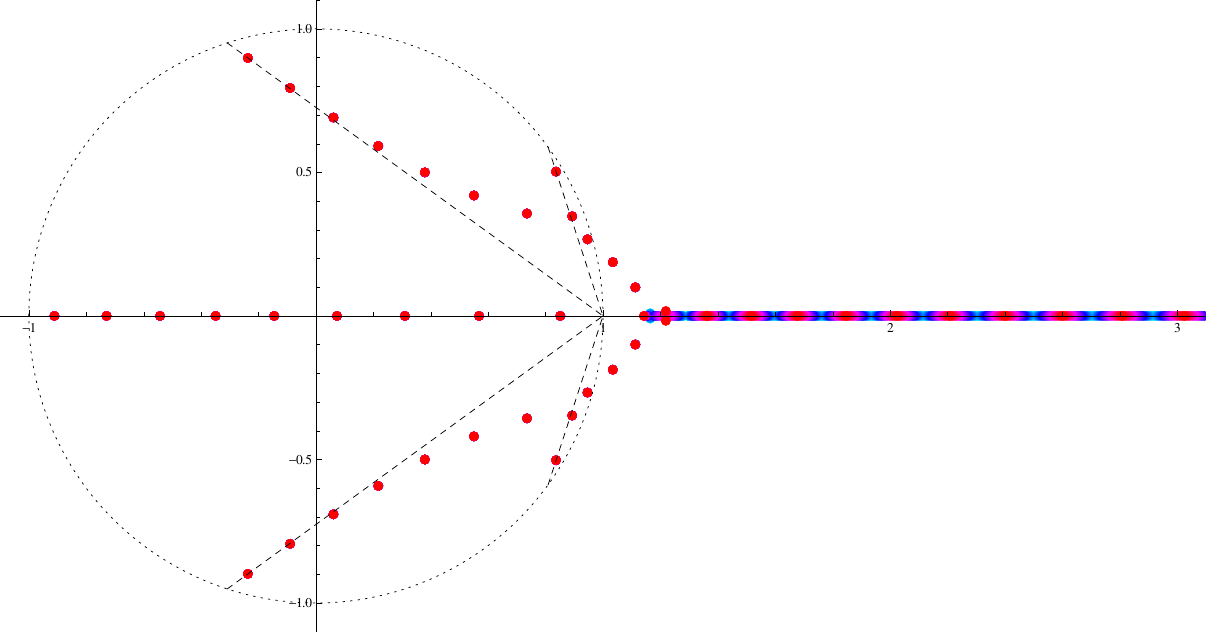}
  \caption{$(4,1)$}
  \label{n_4,1}
  \end{subfigure}

\hspace{0.03\textwidth}
  \begin{subfigure}{2.0cm}
  \includegraphics[width=6cm]{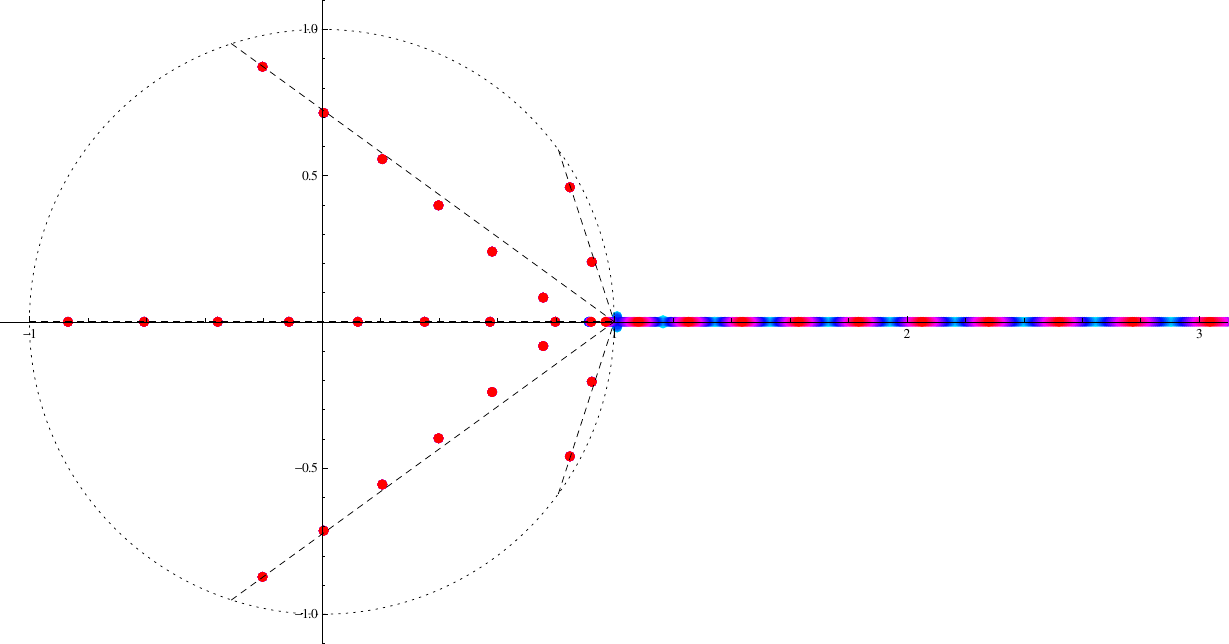}
  \caption{$(3,2)$}
  \label{n_3,2}
  \end{subfigure}
\caption{(Color online) Complex plane of normalized energy $u$, Eq.~(\ref{eq:u}), for $\alpha=200$ and various valences $(n_1,n_2)$. The dotted circle is $|u|=1$, the dashed lines connect spectrum termination points  $u=-(1)^{1/(n_1+n_2)}$ and $u=1$, indicating positions of  narrow complex bands.}
\label{EVinPlane_200}
\end{figure}

Figure \ref{EVinPlane_200} shows normalized spectra for several different combinations of charges on the complex energy plane of $u$, Eq.~(\ref{eq:u}), at large concentration $\alpha=200$. One may notice odd number $n_1+n_2$ or $n_1+n_2-1$ of spectral sequences, consisting of order $\sqrt{\alpha}$ exponentially narrow bands, seen as points. The central sequence goes along the real axis terminating at the bottom of the spectrum near $u=-1$. The other appear in  conjugated pairs terminating near the roots of unity
$u=-(1)^{1/(n_1+n_2)}$. Close to the termination points the band sequences align along the lines pointing towards $u=1$. Further away from the termination points they deviate from these lines and may coalesce.

Although thermodynamics and transport properties of the Coulomb gases are merely determined by the lowest band $\epsilon_0(q)$, below we
address the wider spectral properties of  Hamiltonians (\ref{HamiltonianIso}), presented in Figs.~\ref{EVinPlane_2/1} -- \ref{EVinPlane_200}. To this end
we develop a semiclassical theory which is best suited for the description of exponentially narrow bands present at large concentration  $\alpha\gtrsim 1$.

\section{Monovalent (1,1) gas}
\label{sec:monovalent}
\begin{figure}[h!]
\includegraphics[width=8cm]{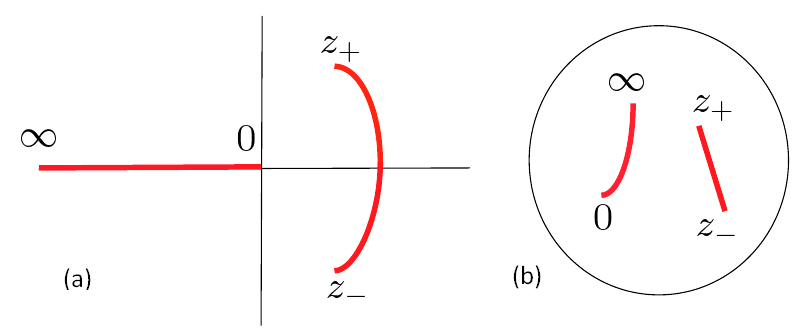}
\caption{ (a) Complex $z$-plane with two cuts. (b) It compactifies to  Riemann sphere  with two  cuts.}
\label{fig:RiemannSphere}
\end{figure}

To introduce the methods, we first develop a semiclassical spectral theory for the Hermitian Hamiltonian (\ref{HamiltonianIso}), (\ref{eq:p}) with $n_1=n_2=1$. To this end we look for wavefunctions in the form $\psi = e^{i\alpha^{1/2}S}$, where $S$ is an action for the classical problem with the normalized Hamiltonian
\beq
                                           \label{eq:cosHam}
2u= p^2-2 \cos \theta \,,
\eeq
where $u=\epsilon/(2\alpha)$, so $u=\mp 1$ correspond to the bottom (top) of the cosine potential.
The semiclassical calculations require knowledge of the action integrals. Our approach to such integrals is based on complex algebraic geometry. First, let $z=e^{i\theta}$ and consider $(z,p)$ as complex variables. Since $p(z)$ resides on the constant energy hypersurface
\beq
                                          \label{eq:EnergySurface}
2u= p^2-\left(z+\frac{1}{z}\right)\,,
\eeq
we have a family of complex algebraic curves
\beq
                                           \label{eq:RiemannSurface}
\mathcal{E}_u:\quad\quad {\cal F}(p,z)=p^2 z - (z^2+2uz+1) = 0
\eeq
parameterized by $u$.  For $u\neq\mp 1$ it can be checked that $(\partial {\cal F}/\partial z,\partial {\cal F}/\partial p)$ does not vanish on ${\cal E}_u$, so each $\mathcal{E}_u$ is nonsingular. Then ${\cal F}(p,z)$ implicitly defines a locally holomorphic map $p=p(z)$. The exceptions to this occur at $z=0,\infty, z_\pm$, where
\beq
                                           \label{eq:TurningPoints}
z_\pm = -u \pm i\sqrt{1 - u^2}
\eeq
are the roots of $p^2=0$ (i.e. classical turning points). In a vicinity of these {\em four} branching points  $p(z)$ behaves as
\begin{align}
&p\sim z^{-1/2},& (z\sim 0)\\
&p\sim z^{1/2},& (z\sim \infty)\\
&p\sim (z-z_\pm)^{1/2},& (z\sim z_\pm)
\end{align}
respectively, i.e. $p(z)$ is locally double-valued. (Note that we have added a point at $z=\infty$ to the complex plane, thereby rendering it compact and topologically equivalent to a Riemann sphere, Fig.~\ref{fig:RiemannSphere}). To make sense of this double-valuedness, we first introduce {\em two} cuts between the four branching points. For convenience we have chosen to do so between $0,\infty$ and the turning points $z_\pm$. Upon this cut domain, $p(z)$ is locally holomorphic.

\begin{figure}[h!]
\includegraphics[width=8cm]{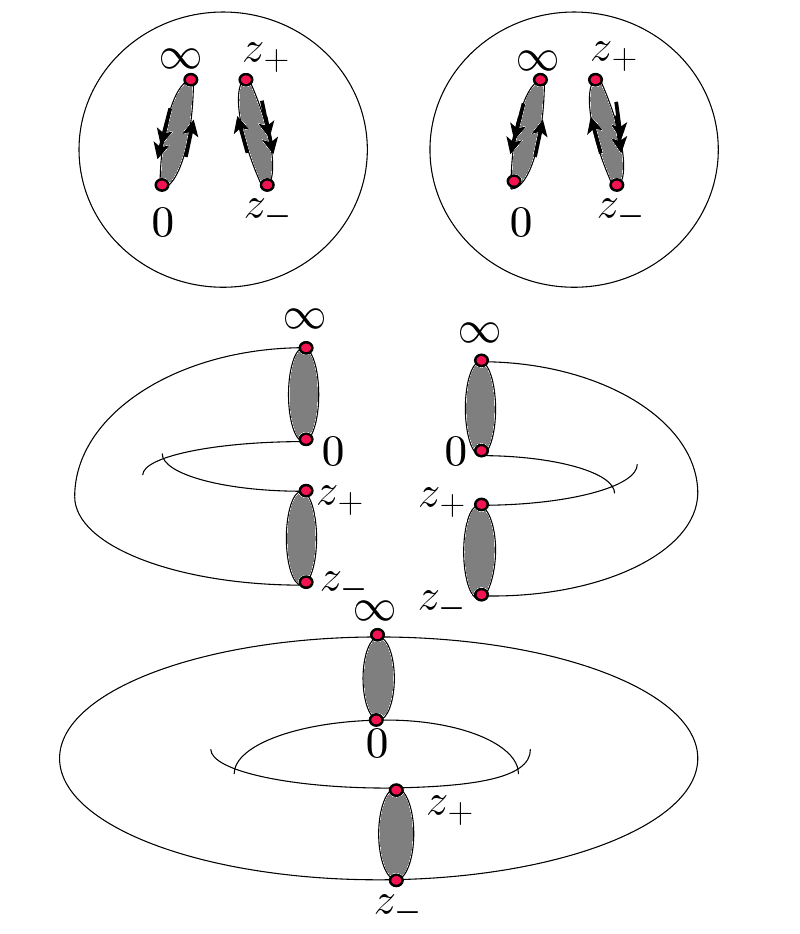}
\caption{Construction of Riemann surface of genus 1. Two Riemann spheres with two cuts each are deformed into tubes to make the gluing in the final step more clear.}
\label{fig:Torus}
\end{figure}

We then introduce a second sheet of the $z$-plane and the corresponding  Riemann sphere, cut in the same way as the first. We then analytically continue $p(z)$ on the first sheet across the cuts onto the second sheet.
If $p(z)$ is analytically continued across the branch cut again, we arrive back on the first sphere where we started. In this way, we obtain $p(z)$ as a locally holomorphic function, whose domain is a doubly-branched cover of the Riemann sphere. Furthermore, suppose we open up the branch cuts, keeping track where on the other branch $p(z)$ will be, if we cross one side of a cut. Identifying these edges one obtains a torus as in Fig.~\ref{fig:Torus} (where the arrows are used to signify the glued together edges). Thus the complex algebraic curve $\mathcal{E}_u$ can be understood as a compact Riemann surface of genus $g=1$ (generically, every compact Riemann surface is topologically a sphere with some number of handles $g$, called the genus of the surface).

\begin{figure}[t!]
\includegraphics[width=8cm]{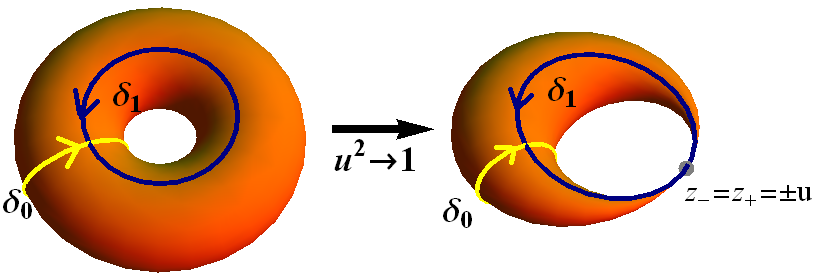}
\caption{Riemann surface of genus-1 with two basic cycles $\delta_0$ and $\delta_1$ on it. In the limit $u\to\mp 1$ the torus degenerates into a singular surface. This coincides with the  loop $\delta_0$ (but not $\delta_1$) becoming contractible to a point.}
\label{fig:DegenerateTorus}
\end{figure}

In the exceptional points $u=\mp 1$ the two turning points collide ($z_+=z_-=\pm1$) and the branch cut between them collapses. The Riemann surface degenerates into a sphere with two points identified, a singular surface of genus 0. This coincides with one of the loops of the torus becoming contractible to a point, Fig.~\ref{fig:DegenerateTorus}. %Note also that these degenerate choices of $u$ correspond precisely to energies coincident with the saddle points of the cosine potential.

\subsection{Integration and topology on torus}

\begin{figure}[h!]
\includegraphics[width=8cm]{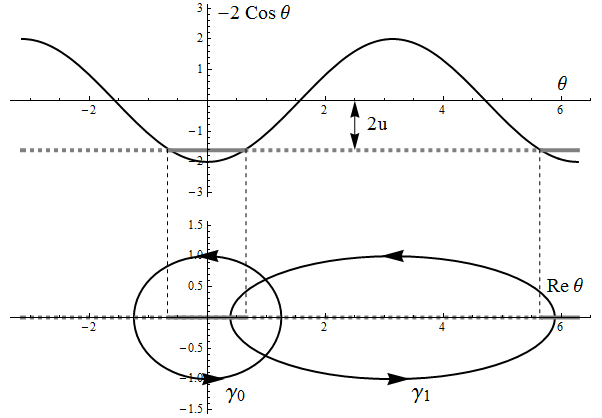}
\caption{The classically allowed (forbidden) region at energy $2u$ are shown by the solid (dashed) gray line.  A classical (instanton) periodic orbit,  in the complex $\theta$-plane, leads to $\gamma_{0} (\gamma_1)$ cycles.}
\label{fig:ThetaPlane}
\end{figure}

The action integrals can be understood  as $S=\oint_\gamma \lambda$ over classical trajectories, where
\beq
                                      \label{eq:ActionForm}
\lambda(u)=p(\theta)\,d\theta=p(z)\frac{dz}{iz}=\frac{(z^2+2uz+1)^{1/2}}{iz^{3/2}}\, dz
\eeq
is the action 1-form which meromorphic on the torus.
To visualize the relevant trajectories we momentarily return to $\theta$ and consider it as complex. In this representation one  has square-root branch cuts along the real axis, connecting the classical turning points. The action integrals run just above or below the real axis in between the turning points. Combining them into  closed cycles, one can push these cycles off the real axis and away from the turning points without altering the action integrals (by Cauchy theorem). The two deformed cycles, shown in Fig.~\ref{fig:ThetaPlane}, are hereafter called $\gamma_0$ and $\gamma_1$.

\begin{figure}[b!]
\includegraphics[width=6cm]{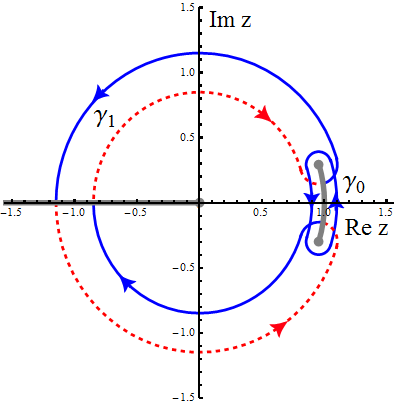}
\caption{(Color online) Cycles $\gamma_0$ and $\gamma_1$ on the complex $z$-plane for $u=-0.9$. Notice that cycle $\gamma_1$ crosses twice the two cuts from first branch (solid blue line) to second branch (dashed red line) and back. }
\label{fig:zPlaneCycles}
\end{figure}

Translating these two cycles to the complex $z$-plane yields the contours of Fig.~\ref{fig:zPlaneCycles}. Notice that these are indeed cycles (i.e. closed contours) owing to the crossing of branch cuts. On the Riemann surface both wind around  the torus. For this reason, the integrals $S_j(u)=\oint_{\gamma_j} \lambda$ are known as \emph{periods} of $\mathcal{E}_u$ with respect to $\lambda(u)$.
One can see that the residue of the action form (\ref{eq:ActionForm}) at infinity is zero. Indeed, at large $z$ we have $\lambda\sim dp$. Therefore we can safely deform the contour around infinity in the $z$-plane.
Let us consider cycles $\delta_0,\delta_1$ as defined in Fig.~\ref{fig:DegenerateTorus}. Any closed cycle on the torus (after appropriate deformation) can be decomposed into a superposition of an integer number of these two basic cycles. For example, the cycles $\gamma_0$ and $\gamma_1$ are
\begin{equation}
                                          \label{eq:HomologyRelation}
\gamma_0=\delta_0,\quad\quad\quad  \gamma_1=2\delta_1-\delta_0\,.
\end{equation}
This is evident if one examines the manner in which these cycles encircle around the torus. Formally, the basic cycles generate the first homology group of the torus (since cycles which are alike in this manner are homologous).

%\emph{This is somewhat misleading, as may be seen by considering a double punctured Riemann sphere. Then there are cycles which can't be contracted to a point, but which yet have no net winding number around either puncture and thus always integrate to zero. But this requires the fundamental group of the surface to be non-abelian, which is never true for a torus. Hence for our purposes one cycle can be deformed into another iff they are homologous.}\\

One can also consider the first \emph{co}homology group of the torus, generated by two independent 1-forms on the Riemann surface modulo exact 1-forms (the latter integrate to zero for all cycles on the torus by Stokes' theorem). In this work we consider meromorphic 1-forms with zero residues. Modulo exact forms they are dual to 1-cycles on the torus by the de Rham theorem\cite{Miranda}. The duality implies that there are exactly as many independent  1-forms to integrate \emph{upon} the surface as independent 1-cycles to integrate \emph{along} the surface. For the torus the cohomology, like the homology, is two-dimensional, i.e. any three (or more) 1-forms on the torus are linearly dependent up to an exact form.

%\emph{The basic thing to establish is that the torus has two-dimesional cohomology. But I'm not sure this is the best way to argue it: Tobias, in his email of 12/12/12, argued that such was a consequence of the Riemann-Roch theorem. However one does it, there must be some amount of 'appeal to mathematical authority' rather than strict demonstration, and it may be better to just emphasize that.}\\

\subsection{Picard-Fuchs equation}
\label{sec:Picard-Fuchs1,1}

As a result, there must exist a linear combination of  1-forms $\{\lambda''(u), \lambda'(u), \lambda(u)\}$ which is an exact form, here primes denote derivatives w.r.t. $u$. This combination may be found by allowing for ($u$-dependent) coefficients in front of the three 1-forms and looking for
an exact form $d_z[P_2(z)z^{-1/2}(z^2+2uz+1)^{-1/2}]$, where $P_2(z)$ is a second degree polynomial with $u$-dependent coefficients. Matching coefficients for powers of $z$ leads to 5 equations for 6 unknown parameters, determining the sought combination up to an overall multiplicative factor. This way one finds that the operator $\mathcal{L}=(u^2-1)\partial_u^2+1/4$ acts on $\lambda(u)$ as
\beq
                                                          \label{eq:exact-form}
\mathcal{L}\lambda(u)=\frac{d}{dz}\left[\frac{i}{2}\, \frac{1-z^2}{z^{1/2}(z^2+2uz+1)^{1/2}}\right]\,.
\eeq %Since $\mathcal{L}$ maps $\lambda$ onto an exact form
It follows from Stokes' theorem and the exactness of $\mathcal{L}\lambda(u)$ that $\mathcal{L}S_j (u)=0$ since $\gamma_j$ is a cycle on the torus. Thus $S_j(u)$ satisfies the linear second order ODE\cite{Gorsky:2010}
\beq
                                     \label{eq:PFeq11}
(u^2-1)S_j^{\prime\prime}(u) + \frac{1}{4}\,S_j(u) = 0\,.
\eeq
This is an example of the Picard-Fuchs equation\cite{Griffiths:1969aa,Deligne:1970aa} (see Ref.~[\onlinecite{Morrison:1991cd}] for a review). Exactly this equation appears extensively in the context of Seiberg-Witten theory.

\begin{figure}[h!]
\includegraphics[width=6cm]{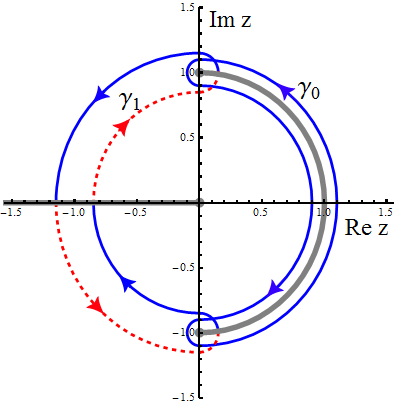}
\caption{The two cycles $\gamma_{0,1}$ for $u=0$. Here $\gamma_1$ may be mapped to $\gamma_0$ by rotating $180^\circ$. }
\label{fig:DiscreteAngle11}
\end{figure}

Inspecting the coefficient in front of the highest derivative, one notices that equation (\ref{eq:PFeq11}) has regular singular points at $u=\infty$ and $u=\mp 1$, where the torus degenerates into a sphere, Fig.~\ref{fig:DegenerateTorus}. Changing variable to $u^2$, this equation may be brought to the standard hypergeometric form\cite{Heckman}. In the  domain $|\arg(1-u^2)|<\pi$ it admits two linearly independent solutions of the form $F_0(u^2)$ and $u F_1(u^2)$, where
\begin{align}
                                                  \label{eq:F0}
F_0(u^2)&= \,_2F_1\left(-\frac{1}{4},-\frac{1}{4};\frac{1}{2};\,u^2\right),\\
F_1(u^2)&= \,_2F_1\left(+\frac{1}{4},+\frac{1}{4};\frac{3}{2};\, u^2\right).\label{eq:F1}
\end{align}
These solutions form a basis out of which $S_j(u)$ (and indeed any period of (\ref{eq:RiemannSurface})) must be composed
\begin{align}
                                                  \label{eq:C}
S_0(u)&=C_{00} F_0(u^2)+C_{01} u F_1(u^2), \\
S_1(u)&=C_{10} F_0(u^2)+C_{11} u F_1(u^2).
                                                 \label{eq:C1}
\end{align}
To find coefficients $C_{jk}$, $j,k=0,1$ appropriate for the action cycles $\gamma_j$ one needs to evaluate the periods at one specific value
of $u$. Employing the fact that the hypergeometric functions (\ref{eq:F0}--\ref{eq:F1}) are normalized and analytic at $u=0$, i.e. $F_k=1+{\cal O}(u^2)$, one notices that $S_j(u)=C_{j0}+u C_{j1}+{\cal O}(u^2)$. Thus to identify $C_{jk}$ we expand $S_j(u)$ to first order in $u$ and evaluate the integrals at $u=0$. The corresponding cycles in the $z$-plane are shown in Fig.~\ref{fig:DiscreteAngle11} and explicit calculation yields
\begin{align}
&C_{00}=e^{-i\pi/2}C_{10}=8\pi^{-1/2}\Gamma(3/4)^2,\\
&C_{01}=e^{+i\pi/2}C_{11}=\pi^{-1/2}\Gamma(1/4)^2.
                                                                         \label{eq:C0}
\end{align}
The relations between $C_{0k}$ and $C_{1k}$ are not accidental. They originate from the fact that for $u=0$ the turning points are $\pm i$ and so the cycle $\gamma_1$ transforms into $\gamma_0$ by substitution $z'=e^{-i\pi}z$, Fig.~\ref{fig:DiscreteAngle11}. Together with Eqs.~(\ref{eq:C}), (\ref{eq:C1}) these relations  imply global symmetry between the two periods
\begin{eqnarray}
                                                                                        \label{eq:DiscreteAngle11}
S_0(u)=e^{-i\pi/2}S_1(e^{i\pi}u)\,.
\end{eqnarray}
Equations (\ref{eq:F0})--(\ref{eq:C0}) fully determine the two actions $S_{0,1}(u)$ through the hypergeometric functions \cite{foot1}. One should now relate them to physical observables.

\subsection{Structure of $S_j(u)$ near $u=-1$}
\label{sec:bottom1,1}

\begin{figure}[h!]
\includegraphics[width=8cm]{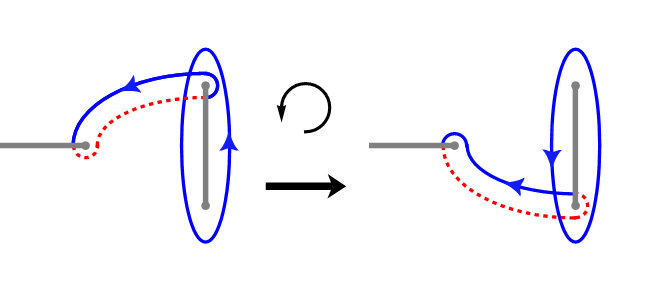}
\caption{(Color online). Monodromy transformation  $(u+1)\to(u+1)e^{2\pi i}$ rotates the branch cut between $[z_-,z_+]$ by $180^\circ$  counter-clockwise.  This changes the cycle $\delta_1\to\delta_1^\prime=\delta_1-\delta_0$ along with it.}
\label{fig:Monodromy11}
\end{figure}

To this end we consider the structure of $S_j(u)$ in the neighborhood of $u=-1$. As noted earlier, the cycle $\gamma_0=\delta_0$  contracts to a point as $u\to- 1$ and therefore $S_0(-1)=0$ by Cauchy's theorem. By contrast, $S_1(-1)$ remains finite. Moreover, while $S_0$ is analytic near $u=-1$, it turns out that $S_1$ is not. To see this, choose some $u\gtrsim -1$ and allow $u$ to wind around $-1$ (i.e. $(u+1)\to(u+1)e^{2\pi i})$. Since $u\approx -1$ the roots $z_\pm$ in $(\ref{eq:TurningPoints})$ are of the form $ z_\pm=-1\pm i\sqrt{2(u+1)}$ we see that this transformation exchanges these branch points via a counter-clockwise half-turn; the branch cut in effect rotates by $180^\circ$.
For the cycle $\delta_0$, which encloses the turning cut, this has no effect: the cut turns within it. Not so for $\delta_1$: as the cut rotates, one must allow $\delta_1$ to continuously deform if $\delta_1$ is never to intersect the branch points.  The overall effect is shown in Fig.~\ref{fig:Monodromy11}. The effect of this \emph{monodromy} transformation is to produce a new cycle $\delta_1^\prime$. Thus, while we have returned to the initial value of $u$, the period $S_1(u)$ (unlike $S_0(u)$) does not return to its original value and so $S_1(u)$ cannot be analytic near $u=-1$.

These facts are consistent, of course, with the origin of the integrals as the classical and instanton actions. At $u\to -1$, the classically allowed region collapses and $p(\theta)\to 0$, so the classical action at the bottom of the cosine potential approaches that of the harmonic oscillator $S_0(u)\propto (1+u)$ (indeed the classical period $T\propto \partial_u S_0$ is a constant). For the instanton trajectory $\gamma_1$ the action $S_1$  does not vanish.  Moreover as $u\to -1$  the period on the instanton trajectory is logarithmically divergent since the trajectory goes to the extrema of the cosine potential, Fig.~\ref{fig:ThetaPlane}. This implies that $S_1(u)\propto {\rm const} + (1+u)\ln(1+u)$.

In fact, more can be said. Under monodromy transformation basis cycle $\delta_1^\prime$ relates to the original basis as $\delta_1^\prime=\delta_1-\delta_0$ (as may be seen by counting intersections of cycles or by moving onto the torus). Thus $(\delta_0,\delta_1)\to(\delta_0,\delta_1-\delta_0)$. From the decomposition of $\gamma_0,\gamma_1$ noted in (\ref{eq:HomologyRelation}) it follows that the $S_j(u)$ must transform as
\beq
                                                                    \label{eq:M-1}
\begin{pmatrix}
S_0(u) \\ S_1(u) \end{pmatrix} \to
\begin{pmatrix} 1 & 0 \\ -2 & 1 \end{pmatrix} \begin{pmatrix} S_0(u) \\
S_1(u) \end{pmatrix} = M_{-1} \begin{pmatrix} S_0(u) \\ S_1(u)
\end{pmatrix} ,
\eeq
where we have introduced the \emph{monodromy matrix} $M_{-1}$ of the actions near $u=-1$.
Since this variation of $S_1$ occurs for every such \emph{monodromy} near $u=-1$, $S_1$ must have a component which depends logarithmically on $1+u$. Indeed, $\ln{(1+u)}$ increases by $2\pi i$ under the monodromy and since $S_1$ changes by $-2S_0$ it must have the following functional form
\beq
                               \label{eq:FuncS111}
S_1(u)=Q_1(u)+\frac{i}{\pi}\, S_0(u)\ln(1+u)\, ,
\eeq
where $Q_1(u)$ and $S_0(u)$ are analytic functions of $(1+u)$.

As an immediate corollary, one can use the relation (\ref{eq:DiscreteAngle11}) between $S_0$ and $S_1$  to find the structure of the solution near $u=+1$. Then the functional form of $S_0(u)$ near $u=+1$ is
%\beq \label{eq:FuncS011}
$S_0(u)=Q_0(u)-iS_1(u)\ln(1-u)/\pi$,
%\eeq
where $Q_0(u)=-i Q_1(-u)$ and $S_1(u)=i S_0(-u)$ are analytic functions of $(1-u)$. The corresponding monodromy matrix is
\beq
                                                         \label{eq:M1}
M_1=\begin{pmatrix} 1 & 2 \\ 0 & 1
\end{pmatrix}.
\eeq

While the structure of the periods near $u=\pm 1$ has been shown through geometric reasoning, it may be also found directly by looking for  solutions
of the Picard-Fuchs equation (\ref{eq:PFeq11}) as  power series in $(1\pm u)$. Such a procedure along with the demand of a constant Wronskian
leads to a realization that one of the two solutions must include $(1\pm u) \ln(1\pm u)$ terms along with the iterative sequence for finding
the coefficients of the polynomials.   This allows for direct verification of Eq.~(\ref{eq:FuncS111}).

\subsection{Semiclassical results}

We now seek semiclassical results for the sequence of low-energy bands terminated at $u=-1$. We shall interpret the period $S_0(u)$ which is analytic around $u=-1$ as a classical action. The latter should be quantized according to the Bohr-Sommerfeld rule to determine the normalized energies $u_m$ of the bands
\beq
                                                  \label{eq:BZ}
S_0(u_m)=2\pi \alpha^{-1/2}(m+1/2)\,,\quad\quad  m=0,1,\ldots
\eeq
(we shall not discuss the origin of the Maslov index $1/2$ here). The second non-analytic period $S_1(u)$ is identified as the instanton action, which determines the bandwidth $(\Delta u)_m$ according to Gamow's formula
\begin{equation}
 \label{eq:Gamow}
 (\Delta u)_m=\frac{\omega}{\pi\sqrt{\alpha}}\, e^{i\alpha^{1/2} S_1(u_m)/2}\,,
\end{equation}
where $\omega=2$ is the classical frequency for the Hamiltonian (\ref{eq:cosHam}). The monodromy of $u$ around $-1$, Eq.~(\ref{eq:M-1}), carries over to the bandwidth as a factor of $e^{(i/2)\alpha^{1/2}(-2S_0(u_m))}$. Then the Bohr-Sommerfeld quantization (\ref{eq:BZ}) is also a condition for the bandwidth to be invariant with respect to monodromies.

To illustrate these results we expand the periods Eqs.~(\ref{eq:C})--(\ref{eq:C0}) near $u=-1$ to find the physical energy levels
$\epsilon_m=2\alpha u_m$. To first order one finds for $S_0(u)$ and $Q_1(u)$
\begin{align}
                                                                    \label{eq:first-order}
S_0(u)&=2\pi(u+1)\, , \\
Q_1(u)&=16i-2i(u+1)\ln{(32e)}\,,
\end{align}
implying $\epsilon_m = -2\alpha+2\alpha^{1/2}\left(m+1/2\right)$.   As a result the pressure (\ref{eq:pressure}) of a monovalent gas is
\begin{equation}
P=-eE_0\epsilon_0 = 2k_BT f - \sqrt{k_BTeE_0 f}.
\end{equation}
The two terms here are the pressure of the ideal gas with the fugacity $f$ and the mean-field Debye-Hueckel interaction correction respectively\cite{ZhangPRE}.

The instanton action, Eq.~(\ref{eq:FuncS111}), at quantized $u_m$ is
\begin{equation}
 \label{eq:instanton}
 S_1(u_m) = 16i+\frac{2i}{\alpha^{1/2}}\left(m+\frac{1}{2}\right)\ln\left(\frac{m+1/2}{32e\alpha^{1/2}}\right)\,,
\end{equation}
where the linear term in $Q_1(u)$ has been absorbed into the logarithm. The Gamow formula (\ref{eq:Gamow}) leads to
\begin{eqnarray}
 \label{eq:bandwidth}
 &&(\Delta\epsilon)_m = 2\alpha(\Delta u)_m = 2\alpha\,\frac{\omega}{\pi\sqrt{\alpha}}\, e^{i\alpha^{1/2} S_1(u_m)/2}\nonumber \\
 &&=\frac{4}{\pi}\left(\frac{32e}{m+1/2}\right)^{m+1/2}\, e^{-8\alpha^{1/2}+(m/2+3/4)\ln\alpha},
\end{eqnarray}
This coincides with the known asymptotic results for the Mathieu equation\cite{CUMS,MeixnerSchaefke,AbramowitzStegun}.
%in reference \onlinecite{CUMS}, where the authors note that it differs from the exact mathematical result only by Stirling's formula \cite{MeixnerSchaefke}.

\subsection{Neighborhood of $u=\infty$}

For completeness we also consider the behavior of the actions at high energy. In the limit $u\to\infty$ the Picard-Fuchs equation (\ref{eq:PFeq11}) is of the form $u^2 S^{\prime\prime}(u)+S(u)/4=0$. Seeking a solution in the form $S=u^r$, one finds $r(r-1)+1/4=(r-1/2)^2=0$ and thus there must be two independent solutions with the leading behavior $u^{1/2}$ and $u^{1/2}\ln(u)$. So the two periods should be of the form
\beq
                                                                     \label{eq:u-infty}
S_i(u)=u^{1/2}\left[V_i(u)+ W_i(u)\ln u \right]\, ,
\eeq
where $W_i, V_i$ are analytic functions of $1/u$. To find these functions one needs to notice that while the continuation to infinity for $S_1$ is unambiguous, the result obtained for $S_0$ depends on whether the path to infinity passes above or below $u=1$. This is due to the fact that $S_0$ exhibits nontrivial monodromy around $u=1$, Eq.~(\ref{eq:M1}). In other words,  whether $u$ goes to infinity below or above the real axis determines which of the two turning points $z_\pm$ goes to zero or infinity. Since these are also branching points for the torus, the path of analytic continuation determines how the cycles on the torus are carried along in the process.

Thus looking for the asymptotic behavior of the periods (\ref{eq:C})--(\ref{eq:C0}) at
$u\to\infty\pm i0$, one finds\cite{AbramowitzStegun}

\begin{align}
V_0(u) &= i\pi W_1(u) \mp V_1(u), \\
W_0(u)&=\mp W_1(u), \\
V_1(u) &= 4i\sqrt{2}\left[ \ln\left(e^2/8\right)+2/u\right], \\
W_1(u) &=-4i\sqrt{2}\left[1-(4u)^{-2} \right],
                                                                     \label{eq:VW}
\end{align}
to leading corrections in $1/u$. Since $S_0(u)\pm S_1(u)=i\pi W_1(u) u^{1/2}$, from here one may readily show that under the monodromy $u\to u e^{2\pi i}$  the two actions transform with the following monodromy matrices
\beq
                                                      \label{eq:MMM}
M_{\infty - i0} =
\begin{pmatrix}
-3 & 2 \\ -2 & 1
\end{pmatrix},\hspace{4mm}
M_{\infty + i0} =
\begin{pmatrix}
1 & 2 \\ -2 & -3
\end{pmatrix}.
\eeq
One may check that the three monodromy matrices satisfy
\beq
M_{\infty-i0}=M_{1} \cdot M_{-1},\hspace{1cm} M_{\infty+i0}=M_{-1} \cdot M_{1},
\eeq
as expected\cite{Heckman}: winding around 0 in large counter-clockwise circle is the same as winding -1 and 1 sequentially counterclockwise.

From Eqs.~(\ref{eq:u-infty})--(\ref{eq:VW}) one finds the unique non-singular period at $u\to\infty\pm i0$ to be given by
$S_0(u)\pm S_1(u)=-i\pi W_1(u) u^{1/2}$. As discussed above, it must be identified with the classical action and  subject
to Bohr-Sommerfeld quantization $(S_0(u_m)\pm S_1(u_m))/2 = 2\pi\alpha^{-1/2} m$. This leads to $u_m\approx m^2/2\alpha$ and thus
$\epsilon_m=2\alpha u_m=m^2$, as expected for the high energy spectrum.

\section{Divalent (2,1) gas}
\label{sec:divalent}

The divalent (2,1) gas is the simplest case where the Hamiltonian (\ref{HamiltonianIso}) is non-Hermitian. Employing complex variable $z=e^{i\theta}$ and normalized energy $u=2\epsilon/3\alpha$, it takes the form
\begin{equation}
                                                          \label{eq:21Ham}
\frac{3}{2}\, u = p^2-\left(\frac{z^2}{2}+\frac{1}{z}\right).
\end{equation}
Similarly to Eq.~(\ref{eq:EnergySurface}) this defines a family of complex algebraic curves
\begin{equation}
\mathcal{E}_u:\quad\quad {\cal F}(p,z)=2 p^2 z - \left(z^3+3 u z+2 \right)=0.
\end{equation}
The map $p=p(z)$ is locally holomorphic away from the zeros $z_0,z_\pm$ (see Fig.~\ref{fig:Paths2,1}). At these three branching points as well as at the singularity at $z=0$ the function $p(z)$ is locally double-valued and behaves as $p\sim (z-z_j)^{1/2},\;j=0,\pm$ and $p\sim z^{-1/2}$, respectively. In contrast to the monovalent $(1,1)$ case, Sec.~\ref{sec:monovalent}, the function $p(z)$ is single-valued at $z\sim\infty$ where it goes as $p\sim z$, so no branch cut extends to $z=\infty$.
Nevertheless there are again four branching points. To construct the Riemann sphere we draw two branch cuts: one between $[0,z_0]$ and the other between $[z_+,z_-]$. The resulting Riemann surface is again $g=1$ torus, analogous to Fig.~\ref{fig:Torus}.

Its moduli space $u$ contains {\em four} singular points $u=-1,e^{\pm i\pi/3}$ and $u=\infty$, where the torus degenerates into the sphere. (There were only three such points in the (1,1) case.) For $u=-1$ the branching points $z_\pm$ coalesce, while for $u=e^{\pm i\pi/3}$
the branching point $z_0$ collides with  $z_\pm$, correspondingly. As $u\to +\infty$, the branching point $z_0$ approaches $z=0$, while $z_\pm \to \pm i\infty$.

\begin{figure}[h!]
 \includegraphics[width=8cm]{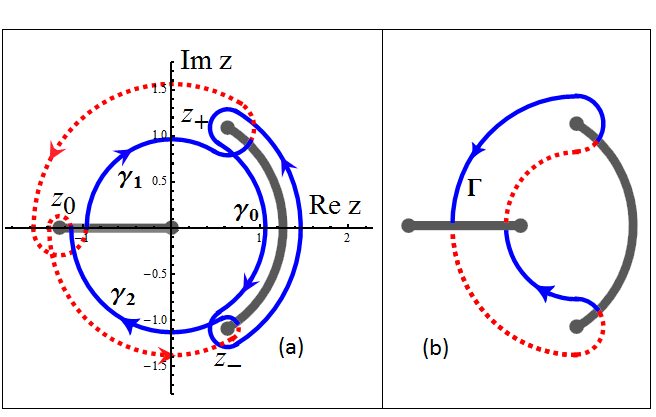}
 \caption{(Color online) Complex $z$-plane with two branch cuts, shown in gray. (a) Three integration cycles $\gamma_0,\gamma_1,\gamma_2$  are displayed for $u=0$. (b) The instanton cycle $\Gamma=-\gamma_1+\gamma_2$. The solid blue (dashed red) lines denote parts of the cycles going over the first (second) branch. }
 \label{fig:Paths2,1}
\end{figure}

The action integrals are again defined  as $S_j=\oint_{\gamma_j} \lambda$, where the 1-form $\lambda(u)=p(z)dz/iz$ is meromorphic on the torus.
In general the counterparts of the turning points in the complex $\theta$-plane are not real. This makes it more convenient to discuss the action cycles $\gamma_j$ in the $z$-plane. With three  turning points $z_0, z_\pm$, it is convenient to take  three paths of integration $\gamma_0,\gamma_1,\gamma_2$, depicted in Fig.~\ref{fig:Paths2,1}. In terms of the two basic cycles on the torus $\delta_0,\delta_1$, Fig.~\ref{fig:DegenerateTorus}, the three paths are given by
\begin{align}
                                                                           \label{eq:HomologyRelation21}
\gamma_0=\delta_0\,, \quad\quad \gamma_1=-\delta_1+\delta_0\,,\quad \quad\gamma_2=\delta_1\,.
\end{align}
One may notice that $\gamma_0-\gamma_1-\gamma_2=0$, and thus $S_0=S_1+S_2$. This equality holds because on a Riemann surface of genus 1 there are only two independent closed cycles. From de Rham's theorem\cite{Miranda} one infers that there are exactly two independent 1-forms. Therefore the three forms  $\{\lambda''(u),\lambda'(u),\lambda(u)\}$ are linearly dependent up to an exact form. Following the root outlined in Sec.~\ref{sec:Picard-Fuchs1,1} (where  $P_2(z)$ is replaced with $P_3(z)$ -- polynomial of degree 3),  one obtains the Picard-Fuchs equation
\begin{equation}
 (u^3+1)S_j''(u)+\frac{u}{4}\,S_j(u)=0\,.
                                                                         \label{eq:PF2,1}
\end{equation}
In agreement with the above discussion, there are regular singular points at the third roots of negative unity, i.e. $u=-1, e^{\pm i\pi/3}$ where the coefficient in front of the highest derivative goes to zero,  and at $u=\infty$.
Two linearly independent solutions $F_0(u^3)$ and $uF_1(u^3)$ of this second-order ODE are given in terms of the hypergeometric functions
\begin{align}
 F_0(u^3)&=\,_2F_1\left(-\frac{1}{6},-\frac{1}{6};\frac{2}{3};-u^3\right)\,,\\
 F_1(u^3)&=\,_2F_1\left(+\frac{1}{6},+\frac{1}{6};\frac{4}{3};-u^3\right)\,.
\end{align}
In this basis the three periods $S_j(u)$, where $j=0,1,2$, are given by
\begin{align}
                                                                \label{eq:C21}
 S_j(u)&=C_{j0}F_0(u^3)+C_{j1}uF_1(u^3)\, .
\end{align}
Since the hypergeometric functions $F_j(u^3\to 0)=1+\mathcal{O}(u^3)$, one notices that  $S_j(u)=C_{j0}+uC_{j1}+\mathcal{O}(u^3)$, as $u\to 0$.
One can thus find constants $C_{jk}$ by explicit evaluation of the actions at $u=0$, i.e. $C_{j0}=S_j(0)$ and $C_{j1}=S_j'(0)$. The corresponding integration paths are shown in Fig.~\ref{fig:Paths2,1} and straightforward integration yields:
\begin{align}
 C_{00}&=C_{10}e^{\pi i/3}=C_{20}e^{-\pi i/3}=\frac{2^{11/6}3\pi^{3/2}}{\Gamma(\frac{1}{6})\Gamma(\frac{1}{3})},\\
 C_{01}&=C_{11}e^{-\pi i/3}=C_{21}e^{\pi i/3}=\frac{3^{1/2}\Gamma(\frac{1}{6})\Gamma(\frac{1}{3})}{2^{11/6}\pi^{1/2}}.
\end{align}
These relations along with Eq.~(\ref{eq:C21}) imply the three-fold symmetry between the actions, cf. Eq.~(\ref{eq:DiscreteAngle11}),
\begin{equation}
 S_0(u)=e^{i\pi/3}S_1\left(e^{-2i\pi/3}u\right)=e^{-i\pi/3}S_2\left(e^{2i\pi/3}u\right)\, .
                                                                  \label{eq:ActionRelation2,1}
\end{equation}

Now one needs to connect the periods (\ref{eq:C21}) with the quantum spectrum. We start by discussing the real branch of the spectrum terminating at the singular point $u=-1$, Fig.~\ref{EVinPlane_2/1}. As $u\to -1$, the two branching points $z_\pm$ coalesce. As a result $\gamma_0$ cycle  degenerates to a point, leading to $S_0(u\to-1)\to 0$,  while $S_{1,2}$ remain finite and actually turn out to be non-analytic. This can be seen by considering the monodromy for a winding of $u$ around $-1$, i.e. $(u+1)\to(u+1)e^{2\pi i}$ (cf. Sec.~\ref{sec:bottom1,1}). Such a transformation exchanges branching points $z_\pm$ by a counter-clockwise $180^\circ$-rotation. This leaves the cycle $\delta_0=\gamma_0$, which encloses these two points, unchanged. On the other hand, the cycle $\delta_1$ picks up a contribution of $-\delta_0$: $\delta_1'=\delta_1-\delta_0$. Thus $\gamma_{1,2}$, Eq.~(\ref{eq:HomologyRelation21}), pick up a contribution of $\pm\delta_0$. As a result, for every monodromy cycle, $S_{1,2}$ pick up a contribution of $\pm S_0$, so locally they are of the form
\begin{equation}
 S_{1,2}(u)=Q_{1,2}(u)\mp\frac{i}{2\pi}S_0(u)\ln(1+u)\,,
                                                                                      \label{eq:monodromyS1_2,1}
\end{equation}
where $Q_{1,2}(u)$ and $S_0(u)$ are analytic functions of $(1+u)$ (moreover $Q_1+Q_2=S_0$, cf. Eq.~(\ref{eq:HomologyRelation21})).
This allows us to identify the period $S_0(u)=(\sqrt{6}\pi/2) (1+u)+{\cal O}((1+u)^2)$ as the classical action, while the instanton action is a combination of the two non-analytic periods $S_{1,2}(u)$.

The corresponding monodromy matrix $M_{-1}$ in e.g. basis $(S_0, S_1)$ (since $S_2=S_0-S_1$ is linearly dependent) is
\begin{equation}
                                                                     \label{eq:M-1_2,1}
\begin{pmatrix}
S_0(u) \\ S_1(u)
\end{pmatrix}
\to
\begin{pmatrix}
1 & 0 \\
1 & 1
\end{pmatrix}
\begin{pmatrix}
S_0(u) \\
S_1(u)
\end{pmatrix}
=M_{-1}
\begin{pmatrix}
S_0(u) \\
S_1(u)
\end{pmatrix}.
\end{equation}
Employing Eqs.~(\ref{eq:HomologyRelation21}), (\ref{eq:ActionRelation2,1}), one finds that at the singular point $e^{i\pi/3}$ ($e^{-i\pi/3}$)  the period $S_{1}(u)$ ($S_2(u)$) is non-singular and goes to zero. It should be thus identified with the classical actions for the branch of the spectrum terminating at the respective singular point, Fig.~\ref{EVinPlane_2/1}. A combination of the remaining two actions $S_0$ and $S_2$ ($S_1$) form the corresponding instanton. The respective monodromy matrices (again in the basis  $(S_0, S_1)$) are found as
\begin{equation}
                                                                     \label{eq:M-2,3_2,1}
M_{e^{i\pi/3}}=
\begin{pmatrix}
1 & -1 \\
0 & 1
\end{pmatrix},
\quad\quad\quad
M_{e^{-i\pi/3}}=
\begin{pmatrix}
2 & -1 \\
1 & 0
\end{pmatrix}.
\end{equation}

\begin{figure}[h!]
 \includegraphics[width=8cm]{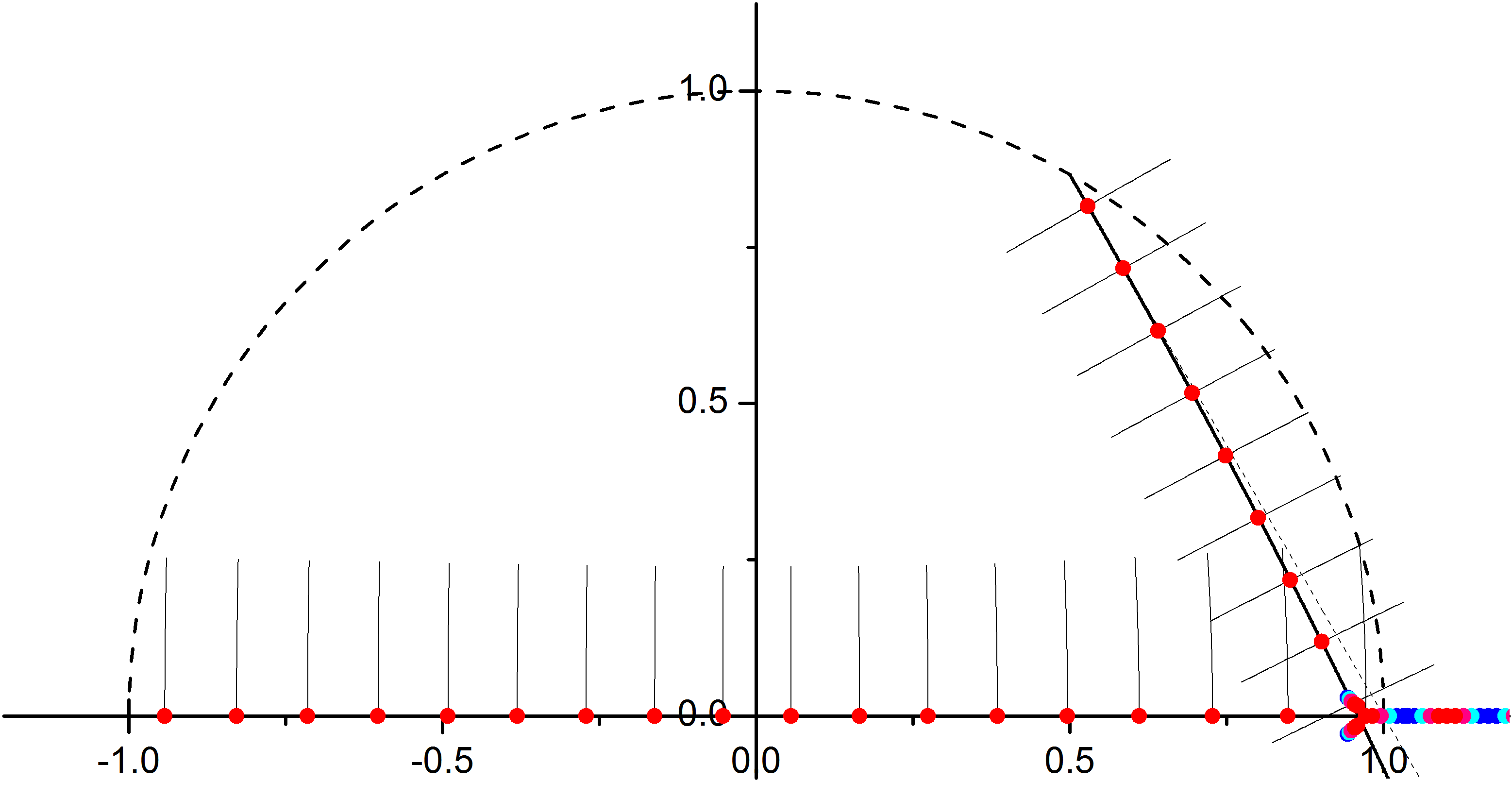}
 \caption{(Color online) Narrow energy bands (red dots) in the upper half-plane of complex energy $u$ for $\alpha=200$, cf. Fig.~\ref{EVinPlane_200}a. $Im\,  S_0(u)=0$ along the real axis, where the small lines mark $Re\,  S_0(u)=2\pi\alpha^{-1/2}(m+1/2)$. The line $Im\,  S_1(u)=0$ emerges from $u=e^{i\pi/3}$ and intersects the real axis at $u\approx0.96$. To the right of this point we observe bands with narrow gaps and use the same coloring convention as in FIGs.~\ref{EVinPlane_2/1}, \ref{EVinPlane_200}. The small perpendicular lines mark $Re\,  S_1(u)=2\pi \alpha^{-1/2}(m+1/2)$.}
 \label{fig:Quantization2,1}
\end{figure}

To find positions of the bands along the three branches of the spectrum, terminating at the three singular points $u=-1, e^{\pm i\pi/3}$, one  employs Bohr-Sommerfeld quantization for the proper classical action $S_j(u)$ with $j=0,1,2$, correspondingly:
\begin{align}
                                                              \label{eq:BZ21}
 S_j(u_m^{(j)})&=2\pi\alpha^{-1/2}(m+1/2), \quad \quad m=0,1,...\, .
\end{align}
Figure \ref{fig:Quantization2,1} shows the lines $Im\,  S_0(u)=0$ and $Im\,  S_1(u)=0$ intersected with the set of lines  $Re\,  S_j(u) = 2\pi\alpha^{-1/2}(m+1/2)$. The numerically computed spectrum sits right at the semiclassical {\em complex} energies $u_m^{(j)}$. The excellent agreement holds all the way up to the point $u\approx 0.96$, where all three periods $S_j$ happen to be purely real. Beyond this point the semiclassical approximation seems to break down, which manifests in e.g. appearance of wide Bloch bands. Expanding $S_0(u)$ near $u=-1$, one finds for the energy levels $\epsilon_m=3u_m^{(0)}\alpha/2$ in the semiclassical approximation $\epsilon_m\approx -3\alpha/2 + \sqrt{6\alpha}(m+1/2)$. The corresponding pressure (\ref{eq:pressure}) $P=-eE_0\epsilon_0$ consists of the two contributions: the ideal $(2,1)$ gas  and the mean-field Debye-Hueckel interaction correction.

\begin{figure}[h!]
 \includegraphics[width=7cm]{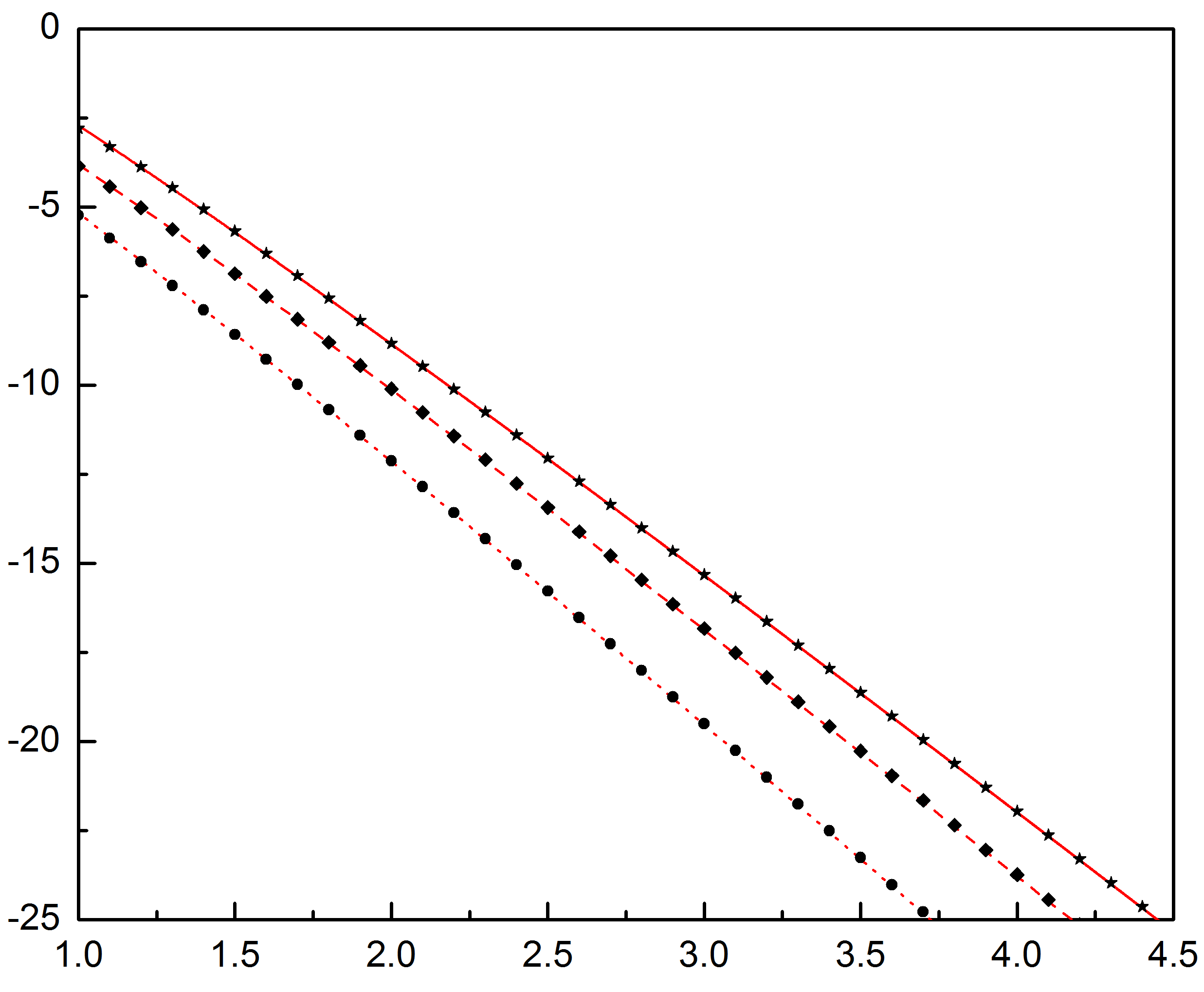}
 \caption{(Color online) Analytic (numerical) results for the logarithm of the bandwidth of the lowest band, $\ln(\Delta\epsilon)_0$, versus square-root of the charge concentration, $\sqrt{\alpha}$, with $(1,1)$ as dotted (circles), $(2,1)$ dashed (diamonds) and $(3,1)$ as solid line (stars).}
 \label{fig:Bandwidth}
\end{figure}

Taking into account that there is no physical difference between $S_1$ and $S_2$ and that the monodromy around $u=-1$ in Eq.~(\ref{eq:monodromyS1_2,1}) should leave the bandwidth in Gamow's formula (\ref{eq:Gamow}) invariant (i.e. it adds a factor of $\exp\{(i/2)\alpha^{1/2}(-2S_0(u_m^{(0)}))\}$), one identifies the instanton cycle with $\Gamma=-\gamma_1+\gamma_2$, Fig.~\ref{fig:Paths2,1}, i.e.  $S_{inst}(u)=-S_1(u)+S_2(u)$. This can be also found by inspecting the cycles in figure \ref{fig:Paths2,1}: one sees that the combined $\Gamma= -\gamma_1+\gamma_2$ cycle connects $z_\pm$ turning points through the ``classically forbidden region'', similarly to $\gamma_1$ instanton cycle in $(1,1)$ case, cf. Fig. \ref{fig:zPlaneCycles}. Note, however, that we do not have a rigorous proof of this fact. Rather our choice of the integration cycle should be considered as an educated guess, which is verified by the numerics.

Expanding $S_{1,2}(u)$ actions near $u=-1$ and substituting $u_m^{(0)}$ from the  Bohr-Sommerfeld quantization (\ref{eq:BZ21}) with $j=0$, one finds for the Bloch bandwidths of the central spectral branch, cf. Eq.~(\ref{eq:Gamow}) with $\omega=\sqrt{6}$,\cite{foot3}
\begin{align}
 (\Delta\epsilon)_m&=\frac{3}{2}\alpha(\Delta u)_m\\
&=\frac{2\sqrt{6}}{\pi}\left(\frac{36\sqrt{6}e}{m+1/2}\right)^{m+1/2} \!e^{-3\sqrt{6\alpha}+(m/2+3/4)\ln\alpha}.\nonumber
\end{align}
Of special interest is the bandwidth of the lowest energy band, due to its direct relation to the transport barrier of the ion channel, Sec.~\ref{Background}. Setting $m=0$ yields
\begin{equation}
 (\Delta\epsilon)_0\approx 34.14\, \alpha^{3/4}\, e^{-7.35 \sqrt{\alpha}}.
 \label{eq:Bandwidth2,1}
\end{equation}
This is in very good agreement with the numerical simulations, Fig.~\ref{fig:Bandwidth}.

Finally we focus on the behavior at $u=\infty$. The Picard-Fuchs equation is of the form $u^3 S^{\prime\prime} +u S/4=0$. Searching for a solution of the form  $S(u)=u^r$ leads to $(r-1/2)^2=0$, signifying  two independent solutions with the leading asymptotic $u^{1/2}$ and $u^{1/2}\ln(u)$. Upon the monodromy transformation $u \to u e^{2\pi i}$ the first of these solutions changes sign, while the second along with the sign change picks up a contribution from the first one. Considering asymptotics of $S_{1,2}(u)$, Eq.~(\ref{eq:C21}), at $u\to +\infty$, one finds the following $SL(2,Z)$ monodromy matrix
\begin{equation}
                                                                     \label{eq:M-infty-2,1}
M_{\infty}=
\begin{pmatrix}
-1 & 0 \\
3 & -1
\end{pmatrix}.
\end{equation}
One can check that
\beq
M_\infty= M_{e^{i\pi/3}}\cdot M_{-1}\cdot M_{e^{-i\pi/3}}\,,
\eeq
as it should be: winding once around 0 in a large counterclockwise rotation is identical to winding counterclockwise in sequence around the other three singular points.

\section{Trivalent (3,1) gas}
\label{sec:trivalent}

%\partial F/\partial p = 0 -> p^2=0 -> z^4+4uz+3=0
%\partial F/\partial z  = 0 -> z^3=-u ->1=z^4->z=1,I,-1,-I -> u=1,-I,1,I

The trivalent (3,1) Hamiltonian with the normalized energy  $u$ is
\beq
                                                                       \label{eq:31Ham}
\frac{4}{3}\, u = p^2-\left(\frac{z^3}{3}+\frac{1}{z}\right)\, .
\eeq
It gives a family of algebraic curves
\beq
\mathcal{E}_u: \quad\quad
{\cal F}(p,z) = 3p^2 z-(z^4+4uz+3)=0
\eeq
over complex $(z,p)$. They are nonsingular if $u^4\neq 1$, and so ${\cal F}(p,z)$ implicitly defines a locally holomorphic map $p=p(z)$ almost everywhere on $(p,z)$. In this case there are {\em six} square-root branching points at $z=0, \infty$ and at the {\em four} turning points, i.e. four roots of $p^2(z)=0$.

\begin{figure}[h!]
\includegraphics[width=8cm]{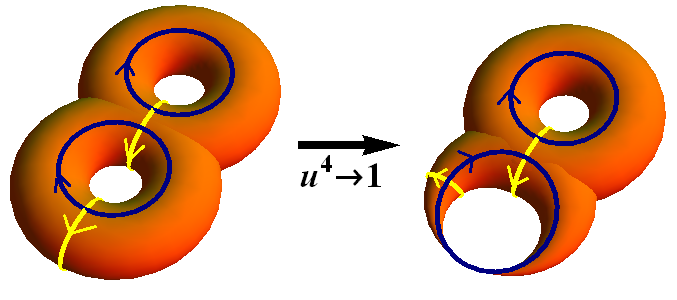}
\caption{(Color online) (a) Double torus curve $\mathcal{E}_u$ for $u^4\neq 1$, having four basic cycles. (b) When $u^4=1$ the $g=2$ torus degenerates into a singular $g=1$ surface. This makes one of the basic cycles to pass through the singularity, and renders another cycle contractible to a point.}
\label{fig:DoubleTorus}
\end{figure}

Hence, while $\mathcal{E}_u$ is a doubly-branched cover of the Riemann sphere, three cuts (instead of two as in the genus-1 case) are required per branch. After opening up cuts and identifying edges under analytic continuation, this leads to a \emph{double torus}, i.e. a sphere with two handles, Fig.~\ref{fig:DoubleTorus}a. Unlike the mono- or di-valent cases, the trivalent channel gives a family of genus-2 Riemann surfaces. The exceptional $u^4=1$ cases  make $\mathcal{E}_u$ singular at $(p,z)=(0,-u)$, due to collision of two turning points, Fig.~\ref{fig:DoubleTorus}b. So the double torus degenerates into a simple torus with two points identified (a singular surface of genus 1).

\begin{figure}[h!]
\includegraphics[width=8cm]{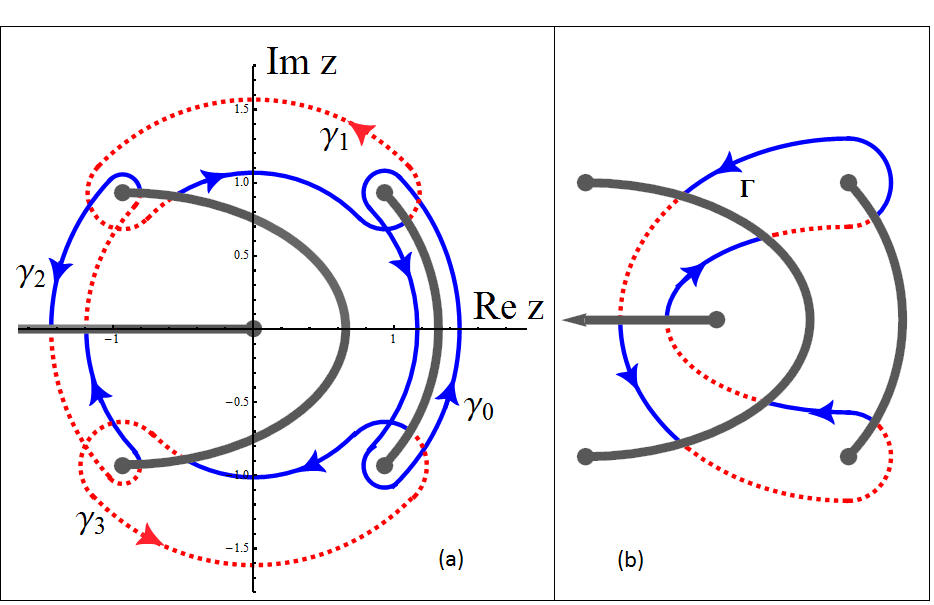}
\caption{(Color online) The Riemann surface is doubly branched with a total of three cuts, shown in gray. The four cycles $\gamma_j$ with $j=0,1,2,3$, along with the instanton cycle $\Gamma$ (defined for later reference) are displayed for $u=0$. The solid blue (dashed red) lines denote parts of the cycles going over the first (second) branch.}
\label{fig:zPlaneCycles31}
\end{figure}

As in the genus-1 cases, the actions can be understood as integrals $S_j=\oint_{\gamma_j} \lambda$ of the meromorphic action 1-form $\lambda(u)=p(z)(dz/iz)$ upon these Riemann surfaces. Owing to the four turning points, there will be four such cycles $\gamma_j$ with $j=0,1,2,3$. These are chosen as in the divalent case, with the inner arcs of each being taken to start on the principal branch. They are shown for $u=0$ in Fig.~\ref{fig:zPlaneCycles31}a. The $u$-dependence of these periods is governed by the Picard-Fuchs equation.

As the double torus is genus-2, there are four independent cycles  (as opposed to two  for genus-1). So the homology---and so too, as argued before, the cohomology---is not two- but {\em four}-dimensional: any {\em five} meromorphic 1-forms on the double torus are linearly dependent up to an exact form. Thus $\lambda(u)$ and its first {\em four} derivatives can be used to produce an exact form; this is done by finding coefficients in a polynomial entering the exact form, as discussed in Sec.~\ref{sec:Picard-Fuchs1,1}. Stokes' theorem  implies that $S(u)=\oint_\gamma \lambda(u)$ must satisfy a 4th-order linear ODE in $u$, i.e. Picard-Fuchs equation which in the present case takes the form
\begin{align}
                                           \label{eq:PF31}
(u^4-1)S^{(4)} +8 u^3 S^{(3)} + \frac{217}{18}\,u^2 S^{\prime\prime} + u S^\prime + \frac{65}{144}\,S = 0.
\end{align}
It has regular singular points at fourth roots of $1$, i.e.  $u\in\{\pm 1,\pm i\}$ and at $u=\infty$.
%and is preserved under the rotation $u\to e^{\pi i/4}u$.
By changing variable to $u^4$, one can cast the  Picard-Fuchs equation as a \emph{generalized hypergeometric equation}. In the cut domain $|\arg(1-u^4)|<\pi$ it has four linearly independent solutions of the form $u^k F_k(u^4)$, where $k=0,1,2,3$ and
\begin{align}
F_0(u^4) =\,_4&F_3\left(-\frac{1}{8},-\frac{1}{8},\frac{5}{24},\frac{13}{24}\,;\,\frac{1}{4},\frac{1}{2},\frac{3}{4}\,; \,u^4\right),\\
F_1(u^4) =\,_4&F_3\left(+\frac{1}{8},+\frac{1}{8},\frac{11}{24},\frac{19}{24}\,;\,\frac{1}{2},\frac{3}{4},\frac{5}{4}\,; \,u^4\right),\\
F_2(u^4)= \,_4&F_3\left(+\frac{3}{8},+\frac{3}{8},\frac{17}{24},\frac{25}{24}\,;\,\frac{3}{4},\frac{5}{4},\frac{3}{2}\,; \,u^4\right),\\
F_3(u^4)= \,_4&F_3\left(+\frac{5}{8},+\frac{5}{8},\frac{23}{24},\frac{31}{24}\,;\,\frac{5}{4},\frac{3}{2},\frac{7}{4}\,; \,u^4\right),
\end{align}
are generalized hypergeometric series. Note that the parameters of each $\,_4F_3(\{a_i\};\{b_j\};u^4)$ satisfy $\sum b_i - \sum a_i = 1$; such hypergeometric series are known as one-balanced or Saalsch\"{u}tzian\cite{Buhring2001}.

Writing the actions in this basis as
\beq
                                             \label{eq:action31}
S_j(u)=\sum_{k=0}^{3}C_{jk} u^k F_k(u^4),
\eeq
we note that $S_j(u)=\sum_{k=0}^{3}C_{jk} u^k+{\cal O}(u^4)$ (as generalized hypergeometric functions are unity at zero and analytic nearby). We expand each $S_j(u)$ up to $u^3$ around $u=0$ and evaluate the resulting integrals, Fig.~\ref{fig:zPlaneCycles31}a, to obtain the $\{C_{jk}\}$ \cite{foot2}. For e.g. $S_0$ this brings
\begin{align}
&C_{00} = +2^{7/2}\cdot3^{-9/8}\pi^{-1/2}\Gamma(5/8)\Gamma(7/8) &\\
&C_{01} = + 2^{-1/2}\cdot 3^{-7/8}\pi^{-1/2} \Gamma(1/8)\Gamma(3/8)& \\
&C_{02} = - 2^{-5/2}\cdot 3^{-13/8}\pi^{-1/2} \Gamma(1/8)\Gamma(3/8)& \\
&C_{03} = - 7\cdot 2^{-1/2}\cdot 3^{-27/8}\pi^{-1/2}\Gamma(5/8)\Gamma(7/8)&
\end{align}
When $u=0$ the turning points satisfy $z^4+3=0$ and so they lie on a certain circle in the complex plane. Hence $\gamma_j$ and $\gamma_{j+1}$ are only different by $\pi/2$ rotation, Fig.~\ref{fig:zPlaneCycles31}a. As a result, we find the four-fold symmetry relations
\begin{equation}
                                     \label{eq:DiscreteAngle31}
S_0(u)=e^{\pi i\over 4}S_1(e^{-{\pi i\over 2}}u)=e^{\pi i\over 2}S_2(e^{-\pi i} u)=e^{-{\pi i\over 4}}S_3(e^{\pi i\over 2} u)
%\nonumber
\end{equation}
for $u$ in the cut domain $|\arg(1-u^4)|<\pi$.

We now consider the periods in the neighborhood of $u=-1$. As before, the cycle $\gamma_0$ becomes contractible to a point as $u\to -1$ and therefore $S_0(-1)=0$ by Cauchy's theorem. The other three actions remain finite, but $S_1$ and $S_3$ are non-analytic. This can be seen by considering the monodromy around $u=-1$. As in the genus-1 cases, the shrinking branch cut near $z=1$  makes a half-turn. Examining the action cycles, it is only $\gamma_1$ and $\gamma_3$ that intersect the cut rotating under the monodromy within the $\gamma_0$ cycle. Hence it is these two cycles that change under monodromy and thus have logarithmic non-analyticity near $u=-1$. More precisely, {$(S_1,S_3)\to (S_1+S_0,S_3-S_0)$} under the monodromy and so these actions are of the form
\begin{align}
                                                       \label{eq:expansion31}
S_{1,3}(u) = Q_{1,3}(u)\mp \frac{i}{2\pi}\,S_0(u)\ln(1+u)\,,
\end{align}
where $Q_{1,3}(u)$ as well as $S_{0}(u)$ and $S_2(u)$ are analytic near $u=-1$.
Since $S_1(u)+S_3(u)$ is seen to be invariant under the monodromy, there are a total of three independent periods which have trivial monodromy around $u=-1$.
This is again supported by considering series solutions of the Picard-Fuchs equation (\ref{eq:PF31}) near $u=-1$. This way one finds three regular solutions with leading behavior $(1+u)^0, (1+u)^1, (1+u)^2$ along with an irregular solution with the leading behavior $(1+u)\ln(1+u)$.
For reasons of space we omit the corresponding $4\times4$ monodromy matrix.

Although analytical facts about $\,_4F_3$ series are sparse (see \cite{Heckman,Buhring2001} for relevant discussion), there are simple consistency checks which our solutions (\ref{eq:action31}) must pass. First the vanishing of the classical action $S_0(u)$ at $u=-1$ implies the identity
\beq
                                \label{eq:FunctionalId31:1}
\sum_{k=0}^3 C_{0k}(-1)^k F_k(1)=0
\eeq
for the hypergeometric functions given above. In addition, from inspection of the Hamiltonian (\ref{eq:31Ham}), one notices that the classical frequency near $u=-1$ is $\omega=\sqrt{8}$. This implies $S_0^{\prime}(-1)=\frac{4}{3}(2\pi/\omega)$ and thus
\beq
                                \label{eq:FunctionalId31:2}
\sum_{k=0}^3 C_{0k} \frac{d}{du}\Big(u^k F_k(u^4)\Big)_{u=-1} = \frac{\sqrt{8}\pi}{3}.
\eeq
Being checked numerically, both hold up to $10^{-16}$.

%Rather than delve further into analysis, it is more fruitful to display the numerical behavior of the $S_i(u)$ functions determined above and compare these with the large-$\alpha$ energy spectra determined from Fourier analysis. Of interest is again Bohr-Sommerfeld quantization, evident in the earlier plot of trivalent energy spectra (Figure \textbf{numerical spectra given earlier in text}). For $\alpha=200$ as in that figure, we plot selected lines of constant real part and imaginary parts for the actions $S_{0,1,3}$ (\textbf{Tobias Figure}). In addition, we superimpose the energy spectra given in \textbf{Figure with numerical spectra}.\\

Now we turn to the analysis of the spectrum of  the Hamiltonian (\ref{eq:31Ham}) at large $\alpha$. There are three spectral branches terminating at
the singular points $u=-1,\pm i$, Fig.~\ref{EVinPlane_200}b (notice that the fourth point $u=1$ lies in the middle of the spectrum and
does not have an obvious semiclassical interpretation). To determine positions of the bands we quantize the corresponding actions $j=0,1,3$ (but not $j=2$, the latter is responsible for the period vanishing at $u=1$) according to the Bohr-Sommerfeld rule:
\begin{equation}
                                                              \label{eq:BZ31}
 S_j(u_m^{(j)})=2\pi\alpha^{-1/2}(m+1/2), \quad m=0,1,...;\, j=0,1,3.
\end{equation}
Figure \ref{fig:Quantization3,1} shows the semiclassical energies $u_m^{(j)}$ along with numerically found energy bands. One notices the perfect agreement
between these two for $Re\,  u \lesssim 1.09$. At the point $u\approx 1.09$ all three actions $S_{0,1,3}$ are purely real and the corresponding instanton action (see below) goes through zero. Beyond this point energy bands are not exponentially narrow and semiclassical approximation may not be applicable.
Notice that this point is unmistakably different from the singular point $u=1$.  Focusing on the real energies at the bottom of the spectrum
and expanding near $u=-1$, one finds with the help of identities~(\ref{eq:FunctionalId31:1}), (\ref{eq:FunctionalId31:2})
$S_0(u)=(\sqrt{8}\pi/3)(1+u) +{\cal O}(1+u)$. The Bohr-Sommerfeld rule (\ref{eq:BZ31}) leads to $\epsilon_m=4u_m^{(0)}\alpha/3= -4\alpha/3 + 2\sqrt{2}\alpha^{1/2}(m+1/2)$. Employing Eq.~(\ref{eq:pressure}), this yields the pressure of the trivalent Coulomb gas as $P = \frac{4}{3}\alpha-\sqrt{2\alpha}$. The two terms here are the ideal gas pressure and the mean-field Debye-Hueckel  correction respectively.

\begin{figure}[h!]
 \includegraphics[width=8cm]{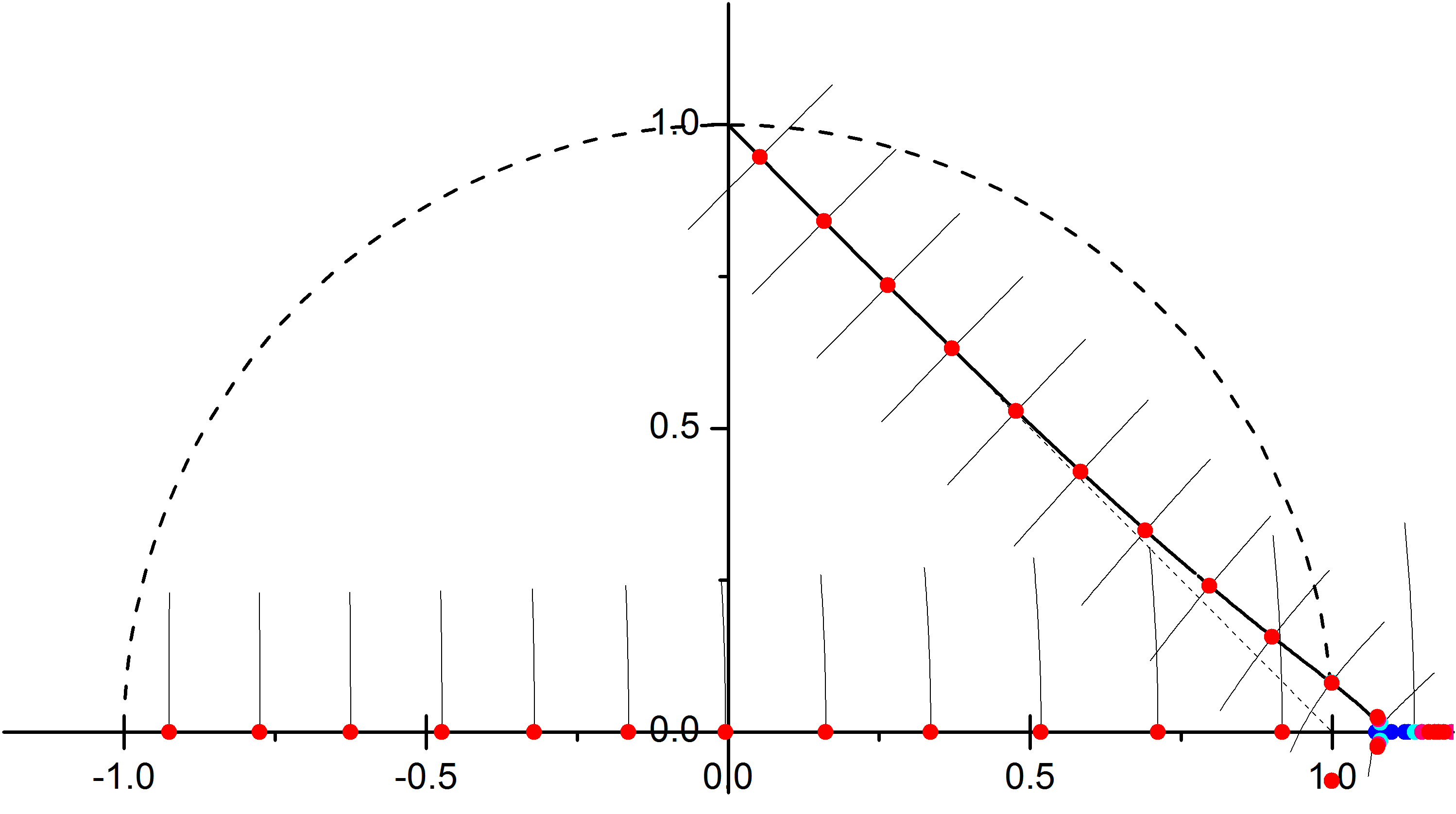}
 \caption{(Color online) Narrow energy bands in the upper half-plane of complex energy $u$ for $\alpha=200$, cf. Fig.~\ref{EVinPlane_200}b. $Im\,  S_0(u)=0$ along the real axis, where the small lines mark $Re\,  S_0(u)=2\pi\alpha^{-1/2}(m+1/2)$. The line $Im\,  S_1(u)=0$ emerges from $u=i$ and intersects the real axis at $u\approx 1.09$. To the right of this point we observe bands with narrow gaps and use the same coloring convention as in FIGs.~\ref{EVinPlane_2/1}, \ref{EVinPlane_200}. The small perpendicular lines mark $Re\,  S_1(u)=2\pi \alpha^{-1/2}(m+1/2)$; red dots, numerically computed narrow bands.}
 \label{fig:Quantization3,1}
\end{figure}

Let us now focus on the width of the Bloch bands near $u=-1$. This requires to identify a cycle corresponding to the instanton action. Guided by the
cosine potential example, cf. Fig.~\ref{fig:ThetaPlane}, we take the corresponding cycle as connecting the turning points of the classical action $S_0$ through the ``classically forbidden region''. This suggests cycle $\Gamma$ shown in Fig.~\ref{fig:zPlaneCycles31}b, which is essentially of the same form as $\gamma_1$ instanton cycle in $(1,1)$ case. One can see that $\Gamma=\gamma_3-\gamma_2-\gamma_1$  by considering intersections of these cycles. Upon the monodromy transformation around $u=-1$ the instanton action thus acquires a contribution $-2S_0(u)$, Eq.~(\ref{eq:expansion31}), which leaves the bandwidth invariant thanks to Bohr-Sommerfeld quantization (\ref{eq:BZ31}). The resulting instanton action is
\beq
                                                                    \label{eq:inst31}
S_{inst}(u)= Q_{inst}(u)+\frac{i}{\pi}S_0(u)\ln(1+u)\,,
\eeq
where $Q_{inst}=Q_3-S_2-Q_1$ is the regular part of $S_{inst}(u)$, cf. Eq.~(\ref{eq:expansion31}). To first order in $(1+u)$ this is ${Q_{inst}(u_m)\approx 14.12i - 6.71i\cdot(1+u)}$, where e.g. the leading term originates from
\begin{align}
&Q_{inst}(-1)=S_{inst}(-1)\nonumber   \\
&=\sum_{k=0}^3\left(C_{3k}-C_{2k}-C_{1k}\right)(-1)^k F_k(1) \nonumber \approx 14.12i.
\end{align}
%(This can be checked numerically by integrating in the $\theta$-plane.)

Then, for $u_m^{(0)}$ along the real $u$-axis satisfying Bohr-Sommerfeld quantization, Gamow's formula yields for the bandwidth
\begin{align}
(\Delta \epsilon)_m=&\frac{4\alpha}{3}(\Delta u) _m=\frac{4\alpha}{3}\cdot \frac{3\omega}{2\pi\sqrt{\alpha}}\, e^{i\alpha^{1/2}S_{inst}(u_m)/2} \\
%=&\frac{4\alpha}{3}\frac{\omega}{\pi\sqrt{\alpha}}\times \left(\frac{\sqrt{2\alpha}/3}{m+1/2}\right)^{m+1/2}\, e^{i\alpha^{1/2}Q_{inst}(u_m)/2}\nonumber \\
\approx&\frac{4\sqrt{2}}{\pi}\left(\frac{582.88}{m+1/2}\right)^{m+1/2}\! e^{-7.06\sqrt{\alpha}+(m/2+3/4)\ln\alpha }.\nonumber
\end{align}
The width of the lowest band $(\Delta \epsilon)_0$ is compared with the numerical results in Fig.\ref{fig:Bandwidth}. As in the earlier cases the two results are in strong accord \cite{foot3}.

For completeness we address the $u=\infty$ behavior. For large $u$ the Picard-Fuchs equation is of the form $u^4 S^{(4)}+8u^3 S^{(3)}+ 217 u^2 S^{\prime\prime}/18 +u S^\prime+65 S/144=0$. The trial  $S(u)=u^r$ brings  four independent solutions with leading asymptotic $\{u^{1/2},u^{1/2}\ln(u),u^{-5/6},u^{-13/6}\}$. The former two are familiar from the genus-1 cases, but the latter two are novel to the genus-2 case. The fractional powers  $\propto 1/6$  may seem unexpected, given the four-fold symmetries of the periods. However, this symmetry is manifest at the level of cycles  at $u=0$, where  four turning points are equally spaced on a circle in the complex $z$-plane. By contrast, as $u\to\infty$, the turning points must satisfy either $z^3\sim -u$ or $1/z\sim -u$, thus only {\em three} of the four turning points tend towards infinity and one towards zero. This leads to the three-fold exchange of actions upon monodromy around $u=\infty$. Thus the $u^r$ behavior of the periods with
$r=-{\rm integer}/(2*3)$ is exactly what is needed to construct a proper $Sp(4,\mathbb{Z})$ monodromy matrix.

\section{Higher valence gases}
\label{sec:32-41}

Here we briefly summarize our current state of understanding of the higher valence $(4,1)$ and $(3,2)$ gases. The corresponding Hamiltonians are
%\partial F/\partial p = 0 -> p^2=0 -> z^4+4uz+3=0
%\partial F/\partial z  = 0 -> z^3=-u ->1=z^4->z=1,I,-1,-I -> u=1,-I,1,IFor completeness, we briefly sketch the state of knowledge for the (4,1)- and (3,2)-valence gases with Hamiltonians
\begin{align}
                                                                       \label{eq:HigherHam}
\text{(4,1)}:\hspace{1.2cm} &\frac{5}{4}\, u = p^2-\left(\frac{z^4}{4}+\frac{1}{z}\right)\,,\\
\text{(3,2)}:\hspace{1.2cm} &\frac{5}{6}\, u = p^2-\left(\frac{z^3}{3}+\frac{1}{2z^2}\right)\,.
\end{align}
In both cases there are {\em five} turning points in the $z$-plane given by the equation $p^2(z)=0$.
The behavior at $z=0$ and $z=\infty$ is somewhat different: for $(4,1)$  there is a branching point at $z=0$, but not at $z=\infty$ (cf. (2,1) problem); while for $(3,2)$ the opposite is true: there is no branching point at $z=0$, but there is one at $z=\infty$. In either case there are {\em six} branching points, which dictate {\em three} branch cuts. The resulting Riemann surface is the double torus, as in $(3,1)$ case, Fig.~\ref{fig:DoubleTorus}. In these cases it is not degenerate as long as $u^5\neq -1$; otherwise two of the five turning points collide, leading to a contraction of one of the cycles.  Therefore one expects five branches of the spectrum terminating at $u=(-1)^{1/5}$, in agreement with Figs.~\ref{EVinPlane_200}c,d.

Since the Riemann surfaces are genus-2, there is a linear combination of the 1-form $\lambda(u)=p(z)dz/iz$ and its {\em four} $u$-derivatives which sum up to an exact form. Therefore any period $S=\oint\!\lambda$ must satisfy a 4th-order ODE in $u$. This is found by matching coefficients in a polynomial entering the exact form (see Sec.~\ref{sec:Picard-Fuchs1,1}), yielding  the Picard-Fuchs equations
\begin{align}
                                                                       \label{eq:PF41}
\text{(4,1):}\hspace{.4cm} &(u^5+1)S^{(4)}(u)+\frac{9u^5-1}{u}\,S^{(3)}(u)\\ \nonumber
 &+\frac{235}{16}\,u^3 S^{\prime\prime}(u) +\frac{5}{4}\,u^2 S^{\prime}(u)+\frac{39}{64}\,u S(u)=0,\\
                                                                       \label{eq:PF32}
\text{(3,2):}\hspace{.4cm} &(u^5+1)S^{(4)}(u)+\frac{9u^5-1}{u}\,S^{(3)}(u)\\ \nonumber
 &+\frac{140}{9}\,u^3 S^{\prime\prime}(u) +\frac{5}{4}\,u^2 S^{\prime}(u)+\frac{119}{144}\,u S(u)=0.
\end{align}\\
While the  coefficients seem arbitrary, some features are notable. First, changing variable to $u^5$, the equations can be brought to the generalized hypergeometric form; one finds four independent solutions of the form $u^k F_k(u^5)$, where $k=0,1,2,4$ and $F_k$ being a certain $_4F_3$ hypergeometric series\cite{foot4}. Notice the absence of a $k=3$ solution. This can be verified directly from the Picard-Fuchs equations, whose leading behavior near $u=0$ is given by $S^{(4)}(u)-u^{-1}S^{(3)}(u)=0$. Substituting $S\propto u^k$, one finds $k(k-1)(k-2)(k-4)=0$.

Second, let us focus on the vicinities of fifth roots of $-1$, e.g. on  $u=-1$. Notably both Eqs.~(\ref{eq:PF41}),(\ref{eq:PF32}) have the same leading behavior $5(u+1)S^{(4)}(u)+10 S^{(3)}(u)=0$, with all other terms are subleading. Looking for a solution in the form  $S(u)\sim (1+u)^s$, one finds for the $s$-exponent $5s(s-1)^2(s-2)=0$. Therefore in both cases there are three analytic solutions with the leading behavior $(1+u)^0,(1+u)^1,(1+u)^2$, while the double root at $s=1$ signifies that the fourth independent solution  is of the form $ (1+u)\ln(1 + u)$\cite{foot5}.

This observation indicates non-trivial monodromy matrix $M_{-1}$,  allowing one to identify the polynomial in front of the $\ln(1+u)$ with the classical action $S_0(u)$. Being quantized according to Bohr-Sommerfeld, the latter determines the spectrum along the branch terminating at $u=-1$, Figs.~\ref{EVinPlane_200}c,d.

Finally, we consider the behavior at $u\to\infty$. By taking trial solutions in the form $S(u)\sim u^r$, one obtains $4$-th order algebraic equations for the exponent $r$. The four roots  of these equations are $\left\{ {1\over 2}, {1\over 2}, -{3\over 4}, -{13\over 4}\right\}$ for $(4,1)$ case and $\left\{ {1\over 2}, {1\over 2}, -{7\over 6}, -{17\over 6}\right\}$ for $(3,2)$ case. Remarkably, there is a double degenerate root at $r=1/2$ in both cases, leading to the two solutions with the leading asymptotic behavior $u^{1/2}$ and $u^{1/2}\ln(u)$. This was also the case in {\em all} the examples, considered above. The first of these solutions, being quantized, leads to $\epsilon_m = m^2$, expected at large energies. The two other roots bring two additional solutions with the leading behavior $u^{-3/4},u^{-13/4}$ or $u^{-7/6},u^{-17/6}$ for $(4,1)$ and $(3,2)$ cases, correspondingly. The denominators of these fractional powers may be related with the fact that {\em four} and {\em three} turning points go to infinity as $u\to \infty$ in the two respective cases. The monodromy transformation $M_\infty$ interchanges the corresponding periods (possibly with a sign change). This is achieved by having  $-{\rm integer}/4$ and $-{\rm integer}/(2*3)$ powers in the corresponding solutions.

%\footnote{That $s=1/2$, if it is a root, must have multiplicity 2 can be shown to be independent of the $S^{\prime\prime}(u)$ and $S(u)$ terms in these Picard-Fuchs equations; requiring that it is such a root implies an additional relation between them (which we shall not state here).},

\section{Connections to Seiberg-Witten Solution}
\label{sec:SW}
Here we briefly review the main features of Seiberg-Witten (SW) solution\cite{Seiberg:1994rs,Seiberg:1994aj}, which were adopted in our calculations\cite{Bilal:1995hc}. The original SW construction gives the spectrum of a four-dimensional supersymmetric $SU(2)$ Yang Mills theory (SYM).
Spectrum of the infrared theory appears to be given by the set of electrically and magnetically charged particles (BPS dyons), which are different from the fundamental particles of the initial UV theory. The latter consists of  a vectormultiplet transforming in the adjoint representation of $SU(2)$, whose components are: one complex scalar field $\phi$, pair of Weyl fermions (gluini) and a $SU(2)$ gauge field (gluon).
In a classical UV vacuum  $\phi$ aligns along the Cartan generator of $\mathfrak{su}(2)$ as $\langle \phi\rangle = a\sigma_3/2$, where the complex expectation value $a$ parameterizes the manifold of classical vacua. In the quantum theory a more convenient coordinate is
\begin{equation}
u=\langle {\rm tr} \phi^2 \rangle\,
\label{eq:udef}
\end{equation}
(such that in the classical limit $u\to \infty$ one has $u\sim a^2$), defining the moduli space of quantum vacua of the theory $\mathcal{M}_u$.

Given the expectation value $a$, one defines the generating function (prepotential) $\mathcal{F}(a)$ as a logarithm of the partition function of the theory, restricted by $\langle \phi\rangle =a\sigma_3/2$. It allows to introduce a canonically conjugated complex variable
\begin{equation}
a_D=\frac{\partial\mathcal{F}(a)}{\partial a}\,,
                                                       \label{eq:prepotential}
\end{equation}
where one may regard $(a, a_D)$ as the coordinate and momentum on $\mathcal{M}_u$.
The underlying supersymmetry allows to argue that $a(u)$ and $a_D(u)$ are holomorphic functions on the
moduli space, safe possibly for few isolated singular points.
In the UV limit $u\to\infty$, one finds a one-loop correction of the form
\begin{equation}
a_D \sim \frac{i a}{\pi}\left(1+ \ln\frac{a^2}{\Lambda^2} \right)\,,
\label{eq:a_Dlargeu}
\end{equation}
where $\Lambda$ is a dynamical scale. Recall that $a\sim\sqrt{u}$ in this region. Therefore, when the argument of $u$ changes by $2\pi i$, $a$ changes its sign and $a_D$ transforms as $a_D\to -a_D+2a$. This rule can be parameterized using the following monodromy matrix in the  $(a_D,a)$ basis
\begin{equation}
M_\infty=\begin{pmatrix} -1 & 2 \\ 0 & -1\end{pmatrix}.
\end{equation}

To find the spectrum of the IR theory means to compute masses of particles which are protected by supersymmetry (so called BPS dyons). BPS mass formula reads
\begin{equation}
M_{n_e, n_m}(u) = |n_e a(u)+n_m a_D(u)|\,,
\label{eq:McentralCharge}
\end{equation}
where $(n_e,n_m)$ are electric and magnetic charges of a dyon respectively, e.g. a monopole has $(n_e,n_m)=(0,\pm 1)$. The above relationship can be understood semiclassically (at large $u$) by evaluating the energy functional for the UV theory on the electrically and magnetically charged configurations. The $\mathcal{N}=2$ supersymmetry guarantees that the very same formula works at strong coupling as well. There are special loci in the $u$ plane where the masses \eqref{eq:McentralCharge} vanish. One can identify these points as singularities for $a$ and $a_D$.

Let us look at the point $u_0$, where the monopole becomes massless $a_D(u_0)=0$. By a conformal transformation one may always scale $u_0=1$. In a vicinity of this point $a_D$ behaves as $a_D\propto (u-1)$, thus near this point $a_D(u)$ is holomorphic, while $a(u)$ is expected to be singular. Performing a one-loop calculation similar to the one near $u=\infty$, in the framework of dual theory, one obtains a relation similar to \eqref{eq:a_Dlargeu}
\begin{equation}
a \sim {ia_D\over \pi}\ln {a_D\over \Lambda}\,.
\label{eq:au1}
\end{equation}
Recalling that $a_D\sim (u-1)$, one finds for the monodromy matrix near $u=1$, again in $(a_D,a)$ basis:
\begin{equation}
M_1=\begin{pmatrix} 1 & 0 \\ -2 & 1\end{pmatrix}.
\end{equation}

From the symmetry considerations one may argue that there should be at least one more singularity in addition to $u=\infty$ and $u=1$. It follows from the fact that if a singularity exists at some value of $u_0$ there ought to be another one at $-u_0$. The $\mathbb{Z}_2$ symmetry, which flips the sign of $u$, is a result of breaking of the global $U(1)$ symmetry (so-called R-symmetry) of IR action. The latter is a remnant of the analogous symmetry in the UV theory which is common for gauge theories with an extended supersymmetry. It exists on the classical level, but is broken by quantum corrections (both perturbative and instanton) down to the $\mathbb{Z}_2$ for $u=\langle {\rm tr}\phi^2 \rangle$.  Therefore, there are at least three singularities in $\mathcal{M}_u$, e.g. at $u=\infty$ and $u=\pm 1$. The third singular point $u=-1$ corresponds to a massless dyon of unit electric and magnetic charges $a(-1)+a_D(-1)=0$. The monodromy matrix around it  can be computed employing  completeness relation  $M_1 M_{-1}=M_\infty$ in the complex $u$-plane.

The non-trivial realization of the SW construction is that complex variables $(a_D(u),a(u))$, with the analytic properties deduced above, may be
viewed as periods of algebraic curves (tori) ${\cal E}_u$, defined over the moduli space ${\cal M}_u$, with respect to some meromorphic differential
$\lambda_{SW}$. The simplest way to parameterize such a curve is
\begin{equation}
{\cal E}_u: \quad  {\cal F}(y,x)= y^2 - (x-u)(x-1)(x+1)=0\,,
                                                                         \label{eq:SWcurvexy}
\end{equation}
where $x,y$ are complex. The above equation describes a double cover of the $x$-plane branched over the four points $x=\pm 1, u$ and $x=\infty$.
Moreover the cover is singular any time two of these points coalesce, i.e at $u=\pm 1,\infty$, as required.
Basis in the first (co)homology of ${\cal E}_u$ (two dimensional in this case) is given by integrals of a one-form over one-cycles. We pick the homology basis $\delta_0, \delta_1$, Fig.~\ref{fig:DegenerateTorus}, and one-form $\lambda_{\text{SW}}(u)$ (SW differential) such that
\begin{equation}
a_D(u)=\int_{\delta_0} \lambda_{\text{SW}}\,,\quad \quad a(u)=\int_{\delta_1} \lambda_{\text{SW}}\,.
                                                                          \label{eq:ABperiods}
\end{equation}
To pick a proper SW differential $\lambda_{\text{SW}}(u)$, we recall that there are only two linearly independent meromorphic 1-forms on the torus up to an exact form.
These two forms may be chosen as  $\lambda_1=dx/y$ and $\lambda_2 =xdx/y$, so $\lambda_{\text{SW}}=\beta_1(u)\frac{dx}{y}+\beta_2(u)\frac{x dx}{y}$, where $\beta_{1,2}(u)$ are functions of $u$ only.
The requirement that the period integrals \eqref{eq:ABperiods} reproduce correct asymptotic behavior of $a(u)$ and $a_D(u)$ at $u=1$ and $u=\infty$ \eqref{eq:a_Dlargeu}, \eqref{eq:au1} allows to determine $\beta_{1,2}(u)$. Finally one obtains
\begin{equation}
\lambda_{\text{SW}}=\frac{\sqrt{2}}{2\pi}\, \frac{\sqrt{x-u}}{\sqrt{x^2-1}}\, dx\,.
\end{equation}
From here one can evaluate the periods (\ref{eq:ABperiods}) in terms of elliptic integrals. They in turn yield the entire information
about BPS mass spectrum  (\ref{eq:McentralCharge}) and the prepotential (\ref{eq:prepotential}).

Close parallels to our calculations are apparent. In fact the SW construction, outlined above, essentially mirrors the $(1,1)$ gas calculations.
The elliptic curve (\ref{eq:SWcurvexy}) is  {\em isogenic} to the torus (\ref{eq:RiemannSurface}) and the two SW periods (\ref{eq:ABperiods}) are directly related to the two action integrals as  $S_0\sim a_D$ and $S_1\sim a+a_D$. In fact, they may be shown\cite{Gorsky:1995zq,Gorsky:2010} to satisfy {\em exactly} the same Picard-Fuchs equation (\ref{eq:PFeq11}) as our actions. Therefore the two basis solutions (\ref{eq:C}), (\ref{eq:C1}), expressible through the complete elliptic integrals of the first and second kind\cite{foot1}, are also a basis for SW periods $a_D(u), a(u)$.

An interesting open question is whether our multivalent examples have analogs in SYM theories. For example, $(2,1)$ case corresponding to a torus with the residual $\mathbb{Z}_3$ symmetry in the $u$-plane, may be related to $SU(2)$ theory with several fundamental hypermultiplets added.
%The SW curve for such theory will change, and one could make a holomorphic change of variables to reduce it to
%the constant energy equation for multivalent gases.
Other examples, leading to $g=2$ surfaces with $\mathbb{Z}_4$ and $\mathbb{Z}_5$ symmetries may be related to certain $SU(3)$ SYM theories with matter.

Another captivating observation is related to the peculiar structure of the spectra near $u\approx 0.96$ in $(2,1)$ gas, $u\approx 1.09$ in $(3,1)$ gas, {\em etc}. These points are marked by the condition $Im\,  S_1(u)/S_0(u)=0$, which is reminiscent of wall crossing phenomena in $\mathcal{N}=2$ theories\cite{Gaiotto:2009hg}. It is observed that moduli space $\mathcal{M}_u$ has domains separated by walls such that when one ``crosses'' a wall the spectrum of the IR theory changes dramatically. For instance, for the $SU(2)$ theory at small $|u|$ there are only two states in the spectrum: monopole $(0,\pm 1)$ and dyon $(\pm 1,\mp1)$. However, at large $|u|$ these particles can form bound states with higher electric charge $(n,\pm 1)$ for any integer $n$. The wall is given by $Im\,  a_D(u)/a(u)=0$.
%The structure of the walls is more sophisticated for theories with higher rank symmetry.

\section{Discussion of the results}
\label{sec:discussion}
In this paper we developed semiclassical treatment for a family of
non-Hermitian ${\cal PT}$-symmetric Hamiltonians. These Hamiltonians
appear
upon transfer-matrix mapping of 1D classical statistical mechanics of
multi-valent Coulomb gases onto quantum mechanics. The low-energy spectra
of the Hamiltonians directly translate into thermodynamic and adiabatic
transport coefficients of the corresponding Coulomb gases.

We use methods of algebraic topology, traditionally employed in the context
of the Seiberg-Witten theory. The main advantage of this strategy is that
it allows us to avoid solving equations of motion and finding classical
trajectories explicitly. The latter task is rather non-trivial (if at all
attainable) in the 4D phase space. Instead, we argue that any surface of
constant energy is a 2D Riemann surface with genus $g\geq 1$. The action along
any closed trajectory (not necessarily satisfying equations of motion) may
be written as an integer valued linear combination of  $2g$ basic periods
of the surface. The latter may be found as solutions of
Picard-Fuchs ODE in the space of parameters. Finally, relations between
basic periods and the quantum spectra are established by considering
special points in the parameter space, where the surface degenerates
into genus $g-1$ singular surface. Consideration of monodromy transformations
in a vicinity of these points allows us to identify classical actions,
quantized according to Bohr-Sommerfeld, as well as the instanton action,
which determines  the bandwidth.

Results obtained this way are in  excellent agreement with numerical
simulations in a broad range of parameters. One of the reasons for this
success is that the method provides with preexponetial factors on
the same footing with the exponent itself.  Another appealing feature of
the approach is that none of our semiclassical calculations required the concept of imaginary
time.
In fact ``time'' (i.e. 1D coordinate of the Coulomb gas) does not appear
at all. In a sense it is substituted by evolution in the space of
parameters of the Hamiltonian (moduli space). We expect the method to be
useful in a broad class of problems which require instanton calculations in
complex spaces.

\section{Acknowledgments}

We are indebted to Alexander Gorsky for introducing us to the
algebraic geometry methods and sharing his unpublished notes. The work was
partially supported by U.S.-Israel Binational Science Foundation Grant
2008075. Research of PK  at the Perimeter Institute is supported by the
Government of Canada through Industry Canada and by the Province of Ontario
through the Ministry of Economic Development and Innovation.


\begin{thebibliography}{99}
\bibitem{foot1} Since the integrals considered here are in fact elliptic integrals over a closed cycle, the hypergeometric functions presented here could have been given directly in terms of the complete elliptic integrals of the first and second kind\cite{Heckman}.
\bibitem{foot3} In writing the Gamow formula above, we conjectured an overall preexponential factor of 4/3 in (2,1) case and 3/2 in (3,1) case compared to the (1,1) case, possibly due to the different structure of the fluctuation determinant. A detailed evaluation of the preexponential factor is beyond the scope of the present work.
\bibitem{foot2} Note that the integrals that arise at $u^2$-order and higher are divergent near the turning points; however, they are convergent near 0 and $\infty$ and can be calculated by deforming the contours to run between these points.
\bibitem{foot4} While we omit the parameters of these series for reasons of space, we note that they satisfy the one-balanced condition\cite{Buhring2001} stated in the (3,1) case.
\bibitem{foot5} Existence of $3=4-1$ analytic solutions near $u=-1$ follows from a theorem of generalized hypergeometric equations going back to Pochammer\cite{Heckman}; the analogous behavior of the mono-, di-, and trivalent gases near $u=-1$ are also instances of this theorem.
\end{thebibliography}
\end{document}